\documentclass[useAMS,usenatbib]{mn2e}

\usepackage{threeparttable}
\usepackage{bigints}
\usepackage{bm}
\usepackage{mdwlist}
\usepackage[section]{placeins}
\usepackage{placeins}
\usepackage{sidecap}
\usepackage{array}
\usepackage{subfigure}
\usepackage{subfig}
\usepackage{float}
\usepackage{caption}

\usepackage{booktabs}
\usepackage{comment}
\usepackage{deluxetable}

\usepackage{natbib}
\usepackage{amssymb,amsmath,amsfonts}
\usepackage{epsfig}
\usepackage{mathptmx}
\usepackage{longtable}
\usepackage{xspace}
\usepackage{graphicx}
\usepackage{tabularx}
\usepackage{multicol}
\usepackage{color}
\usepackage{auto-pst-pdf}

\def\reff@jnl#1{{\rm#1\/}}

\def\aj{\reff@jnl{AJ}}                  
\def\araa{\reff@jnl{ARA\&A}}            
\def\apj{\reff@jnl{ApJ}}                        
\def\apjl{\reff@jnl{ApJ}}               
\def\apjs{\reff@jnl{ApJS}}              
\def\ao{\reff@jnl{Appl.Optics}}         
\def\apss{\reff@jnl{Ap\&SS}}            
\def\aap{\reff@jnl{A\&A}}               
\def\aapr{\reff@jnl{A\&A~Rev.}}         
\def\aaps{\reff@jnl{A\&AS}}             
\def\azh{\reff@jnl{AZh}}                        
\def\baas{\reff@jnl{BAAS}}              
\def\jrasc{\reff@jnl{JRASC}}            
\def\memras{\reff@jnl{MmRAS}}           
\def\mnras{\reff@jnl{MNRAS}}            
\def\pra{\reff@jnl{Phys. Rev. A}}         
\def\prb{\reff@jnl{Phys. Rev. B}}         
\def\prc{\reff@jnl{Phys. Rev. C}}         
\def\prd{\reff@jnl{Phys. Rev. D}}         
\def\prl{\reff@jnl{Phys. Rev. Lett}}      
\def\pasp{\reff@jnl{PASP}}              
\def\pasj{\reff@jnl{PASJ}}              
\def\qjras{\reff@jnl{QJRAS}}            
\def\skytel{\reff@jnl{S\&T}}            
\def\solphys{\reff@jnl{Solar~Phys.}}    
\def\sovast{\reff@jnl{Soviet~Ast.}}     
\def\ssr{\reff@jnl{Space~Sci.Rev.}}     
\def\zap{\reff@jnl{ZAp}}                        
\def\nat{\reff@jnl{Nature}}             

\def\p#1by#2{{\partial{#1} \over \partial{#2}}}
\def\pp#1by#2#3{{\partial^2{#1} \over \partial{#2}\partial{#3}}}
\def\d#1by#2{{{\rm d}{#1} \over {\rm d}{#2}}}
\def\dd#1by#2#3{{{\rm d}^2{#1} \over {\rm d}{#2}{\rm d}{#3}}}

\graphicspath{{./}}

\title[{\sc{CARMA}} 31-GHz follow-up of massive galaxy clusters at $z\gtrsim 0.5$]{{\sc{CARMA}} observations of massive {\it{Planck}}-discovered cluster candidates at $z \gtrsim 0.5$ }
\title[{\sc{CARMA}} follow-up of candidate high-z, massive galaxy clusters]{{\sc{CARMA}} observations of massive {\it{Planck}}-discovered cluster candidates at $z \gtrsim 0.5$  associated with WISE overdensities:  Breaking the size-flux degeneracy}

\label{firstpage}
\author[Rodr{\'i}guez-Gonz{\'a}lvez et al.]{C. Rodr{\'i}guez-Gonz{\'a}lvez$^1$,
R. R. Chary$^1$,
S. Muchovej$^2$,
J.-B. Melin$^3$,
F. Feroz$^4$,
M. Olamaie$^4$,\\
\newauthor{T. Shimwell$^5$}
 \vspace{0.03in}\\
 $^1$ U.S. Planck Data Center, MS220-6, Pasadena, CA 91125, USA, \\
 $^2$ Owens Valley Radio Observatory, California Institute of Technology, Big Pine, CA 93513, USA\\
 $^3$ DSM/Irfu/SPP, CEA-Saclay, F-91191 Gif-sur-Yvette Cedex, France\\
 $^4$ Astrophysics Group, Cavendish Laboratory, JJ Thomson Avenue, Cambridge CB3 0HE, UK \\
 $^5$ Leiden Observatory, Leiden University, 2300 RA Leiden, the Netherlands.\\
 }

\begin{document}
\maketitle

\begin{abstract}
We use a Bayesian software package to analyze {\sc{CARMA}}-8 data towards 19 unconfirmed {\it{Planck}} SZ-cluster candidates from \cite{rodriguez2014a} that are associated with significant overdensities in {\sc{WISE}}. We used two cluster parameterizations, one based on a (fixed shape) generalized-NFW pressure profile and another based on a $\beta$ gas density profile (with varying shape parameters) to obtain parameter estimates for the nine CARMA-8 SZ-detected clusters. We find our sample is comprised of massive, $\left<Y_{500}\right>=0.0010 \pm 0.0015$\,$\rm{arcmin}^2$, relatively compact, $\left<\theta_{500}\right>= 3.9 \pm 2.0\arcmin$ systems. Results from the $\beta$ model show that our cluster candidates exhibit a heterogeneous set of brightness-temperature profiles.
Comparison of {\it{Planck}} and CARMA-8 measurements showed good agreement in $Y_{500}$ and an absence of obvious biases. We estimated the total cluster mass $M_{500}$ as a function of $z$ for one of the systems; at the preferred photometric redshift of 0.5, the derived mass, $M_{500} \approx 0.8 \pm 0.2 \times 10^{15}M_{\odot}$. Spectroscopic Keck/MOSFIRE data confirmed a galaxy member of one of our cluster candidates to be at $z=0.565$. Applying a {\it{Planck}} prior in $Y_{500}$ to the CARMA-8 results reduces uncertainties for both parameters by a factor $>4$, relative to the independent {\it{Planck}} or CARMA-8 measurements. We here demonstrate a powerful technique to find massive clusters at intermediate ($z \gtrsim 0.5$) redshifts using a cross-correlation between {\it{Planck}} and WISE data, with high-resolution follow-up with CARMA-8. We also use the combined capabilities of {\it{Planck}} and CARMA-8 to obtain a dramatic reduction by a factor of several, in parameter uncertainties.
\end{abstract}

\section{Introduction}

The {\it{Planck}} satellite (\citealt{tauber2010} and \citealt{planck1}) is a third generation space-based mission to study the Cosmic Microwave Background (CMB) and its foregrounds. It has mapped the entire sky at nine frequencies from 30 to 857\,GHz,  with an angular resolution of $33\arcmin$ to 5\arcmin, respectively. Massive clusters have been detected in the {\it{Planck}} data  via the Sunyaev-Zel'dovich (SZ) effect \citep{Sunyaev_1972}. {\it{Planck}} has published a cluster catalog containing 1227 entries, out of which 861 are confirmed associations with clusters. 178 of these were previously unknown clusters
while a further 366 remain unconfirmed (\citealt{Planck2013}). The number of cluster candidates identified
in the second data releases, PR2, has now reached 1653. Eventually, the cluster counts will be used to measure the cluster mass
function and constrain cosmological parameters \citep{PlanckCosmos}. However, using cluster counts to constrain cosmology relies, amongst other things, on understanding the completeness of the survey and measuring both the cluster masses and redshifts accurately (for a comprehensive review see, e.g., \citealt{voit2005} and \citealt{allen2011}). To do so, it is crucial to identify sources of bias and to minimize uncertainty in the translation from cluster observable to mass. Regarding cluster mass, since it is not a direct observable, the best mass-observable relations need to be characterized in order to translate the {\it{Planck}} SZ signal into a cluster mass.

The accuracy of the {\it{Planck}} measurements of the integrated SZ effect at intermediate redshifts where, e.g., X-ray data commonly reach out to, is limited by its resolution ($\approx10\arcmin$\, at SZ-relevant frequencies) because the integrated SZ signal exhibits a well-known degeneracy with the cluster angular extent (see e.g., \citealt{Planck2013}). Higher resolution SZ follow-up of {\it{Planck}}-detected clusters can help constrain the cluster size by measuring the spatial profile of the temperature decrement and identify sources of bias. Moreover, a recent comparison of the integrated SZ signal measured by the Arcminute MicroKelvin Imager (AMI; \citealt{zwart2008}) on arcminute scales and by {\it{Planck}} showed that the {\it{Planck}} measurements were systematically higher by $\approx 35\%$ (\citealt{PlanckAMI}). This study, and its follow-up paper on 99 clusters \citep{perrott2014}, together with another one by \cite{muchovej2012} comparing CARMA-8 and {\it{Planck}} data towards two systems, have demonstrated that cluster
parameter uncertainties can be greatly reduced by combining both  datasets.

\begin{table*}
\caption{Details of the sample of {\it{Planck}}-detected cluster candidates analyzed in this work and their CARMA-8 observations; for additional information, the reader is strongly encouraged to look at Paper 1. For simplicity and homogeneity in the cluster naming convention we use a shorthand ID for the targets. The PSZ (Union catalog) name (Planck Collaboration 2013 XXIX) is provided, where available$^{\dag}$. The Right
Ascension (RA) and Declination (Dec) coordinates (J2000) correspond to the map centre of our observations while the cluster coordinates are in Table 4. For the short and long baseline data we provided the visibility noise. Targets that have been detected in the CARMA-8 data have their ID highlighted. The SZ decrement towards P014 had a low signal-to-noise-ratio (SNR) (4.2) and the SZ signal in the CARMA-8 imaged data was considered tentative in Paper 1.}
 \label{tab:clusters}
 \tabcolsep=0.13cm
\begin{tabular}{lcccccc}
\hline \hline
Cluster  & Union Name  & RA & Dec  & Short Baseline (0-2k$\lambda$)   & Long Baseline (2-8 k$\lambda$)    \\
ID               &                          &       &           & $\sigma$                                            &$\sigma$\\
               &                          & hh mm ss  & deg min sec          &    $(\rm{mJy/beam})^{\rm{b}}$  &    $(\rm{mJy/beam})^{\rm{b}}$                    \\ \hline

$\bf{P014}	$ & PSZ1G014.13+38.38 & 16 03 21.62 & 03 19 12.00    	 & 0.309 	 	 & 0.324\\
P028	 & PSZ1G028.66+50.16 & 15 40 10.15 & 17 54 25.14   	  	 & 0.433 	 	 & 0.451\\
P031	 & - & 15 27 37.83 & 20 40 44.28    	  	 & 0.727 	 	 & 0.633\\
P049	 &  - &  14 44 21.61 & 31 14 59.88    	  	 & 0.557 	 	 & 0.572\\
P052	 &  - & 21 19 02.42 & 00 33 00.00    	  	 & 0.368 	 & 0.386\\
P057	 & PSZ1G057.71+51.56 &15 48 34.13 & 36 07 53.86    	 & 0.451 	  & 0.482\\
$\bf{P086}	$ & PSZ1G086.93+53.18 & 15 13 53.36 & 52 46 41.56   	  	 & 0.622 	 	 & 0.599\\
P090	 & PSZ1G090.82+44.13 &16 03 43.65 & 59 11 59.61   		 & 0.389 	 	 & 0.427\\
$\bf{P097} $&  - & 14 55 13.99 & 58 51 42.44   	   & 0.653 	 	 & 0.660\\
$\bf{P109}$ & PSZ1G109.88+27.94 & 18 23 00.19 & 78 21 52.19   	  	 & 0.562  	 & 0.517\\
P121	 & PSZ1G121.15+49.64 & 13 03 26.20 & 67 25 46.70   	 	 & 0.824 	 & 0.681\\
P134	 & PSZ1G134.59+53.41  & 11 51 21.62 & 62 21 00.18   	 	 & 0.590 	 & 0.592\\
P138	 & PSZ1G138.11+42.03 & 10 27 59.07 & 70 35 19.51    		 & 2.170    & 0.982\\
$\bf{P170}$ & PSZ1G171.01+39.44	 & 08 51 05.10 & 48 30 18.14  		 & 0.422 	 & 0.469\\
$\bf{P187}$ & PSZ1G187.53+21.92	 & 07 32 18.01 & 31 38 39.03 	  	 & 0.411 	 & 0.412\\
$\bf{P190}	$& PSZ1G190.68+66.46 & 11 06 04.09 & 33 33 45.23	 	 & 0.450 	 & 0.356\\
$\bf{P205}$ & PSZ1G205.85+73.77	 & 11 38 13.47 & 27 55 05.62 	  	 & 0.385 		 & 0.431\\
P264	 & - &  10 44 48.19 & -17 31 53.90  	 	 & 0.476 	 & 0.513\\
$\bf{P351}	$ & - &  15 04 04.90 & -06 07 15.25		 & 0.355 	  	 & 0.392\\ \hline

 \end{tabular}
 
 \begin{tablenotes}
 \scriptsize
 \item ($\dag$)  Since the cluster selection criteria, as well as the data for the cluster extraction, are different to those for the PSZ catalog, not all the clusters in this work have an official {\sc{Planck}} ID.
 \item (a) Achieved rms noise in corresponding maps.

  \end{tablenotes}
\end{table*} 

In this work, we have used the eight element CARMA interferometer, CARMA-8 (see  \citealt{muchovej2007} for further details) to undertake high spatial resolution follow-up observations at 31 GHz towards 19 unconfirmed {\it{Planck}} cluster candidates\footnote{The {\it{Planck}} SZ catalog used for the initial selection was an intermediate {\it{Planck}} data product known internally as DX7. {\it{Planck}} data are collected and reduced in blocks of time. The DX7 maps used in this analysis correspond to the reduction of {\it{Planck}} data collected from 12$^{\rm{th}}$ August 2009 to the 28$^{\rm{th}}$ of November 2010, which is the equivalent to 3 full all-sky surveys, using the v4.1 processing pipeline. The DX7 maps used in this work are part of an internal release amongst the {\it{Planck}} Collaboration members and, thus, is not a publicly available data product. It should be noted that the public DR1 PSZ catalog supersedes the preliminary DX7 catalog used
for our selection.} (Table \ref{tab:clusters}). Our primary goal was to attempt to identify massive clusters at high redshifts. For this reason, our candidate clusters were those {\it Planck} SZ-candidates that had significant overdensities of galaxies in the {\sc{WISE}} early data release \citep{wright2010} ($\gtrsim$1 galaxy/arcmin$^2$) and a red object\footnote{We describe as {\it{red}}, objects whose  [3.4]-[4.6] {\sc{WISE}} colors are $> -0.1$ (in AB mags, 0.5 in Vega). This is known as the Mid-InfraRed (MIR) criterion and has been shown by e.g., \cite{papovich2008} to preferentially select $z>1$ objects.} within $2.5\arcmin$ fainter than 15.8 Vega magnitudes in the {\sc{WISE}} 3.4-micron band  (which corresponds to a 10$\rm{L}_{*}$ galaxy at $z\approx 1$\footnote{WISE is sensitive to a galaxy mass of $5\times10^{11}$\,$\rm{M}_{\odot}$ at $z\approx 1$. }), to maximize the chances of choosing $z>1$ systems. Similar work using {\sc{WISE}} to find distant clusters has been undertaken by the Massive and Distant Clusters of WISE Survey (MaDCoWS), which in \cite{gettings2012} confirmed their first $z\approx 1$ cluster.

This work is presented as a series of two articles. The first one, \cite{rodriguez2014a}, henceforth Paper 1, focused on the sample selection, data reduction, validation using ancillary data and photometric-redshift estimation. This second paper is organized as follows. In Section \ref{sec:analysis} we describe the cluster parameterizations for the analysis of the CARMA-8 data and present cluster parameter constraints for each model. In addition, we include Bayesian Evidence values between a model with a cluster signal and a model without a cluster signal to assess the quality of the detection and identify systems likely to be spurious. {\it{Planck}}-derived cluster parameters and estimates of the amount of radio-source contamination to the {\it{Planck}} signal are given in Section \ref{sec:planck}. Improved constraints in the $Y_{500}-\theta_{500}$ plane from the application of a {\it{Planck}} prior on $Y_{500}$ to the CARMA-8 results are provided in Section \ref{sec:joint}. In Section \ref{sec:disc} we discuss the properties of the ensemble of cluster candidates, including their location, morphology and cluster-mass estimates and present spectroscopic confirmation for one of our targets. In this section we also 
compare the {\it{Planck}} and CARMA-8 data and  show how our results relate to similar studies.
We note that, for homogeneity, since not all the cluster candidates in this work are included in the PSZ (Union catalog; Planck Collaboration 2013 XXIX), we assign a shorthand cluster ID to each system (see Table \ref{tab:clusters}).

Throughout this work we use J2000 coordinates, as well as a $\Lambda$CDM cosmology with $\Omega_{\rm{m}}=0.3$, $\Omega_{\Lambda}=0.7$, $\Omega_{\rm{k}}=0$, $\Omega_{\rm{b}}=0.041$, $w_o=-1$, $w_{\rm{a}}=0$ and $\sigma_8=0.8$. $H_0$ is taken as 70\,km\,s$^{-1}$\,Mpc$^{-1}$.
 
\section{Quantitative Analysis of CARMA-8 Data}
\label{sec:analysis}

\subsection{Parameter Estimation using Interferometric Data}

In this work we have used {\sc{McAdam}}, a Bayesian analysis package, for the quantitative analysis of the cluster parameters. 
 This package has been used extensively to analyze cluster signals in interferometric data from AMI (see e.g., \citealt{michel2012},  \citealt{carmen2012} \& \citealt{xmm2013} for real data and \citealt{olamaie2012} for simulated data) and once before on CARMA-8 data \citep{shimwell2013}.
{\sc{McAdam}} was originally developed by \cite{marshall2003} and later adapted by \cite{feroz_2009} to work on interferometric SZ data using an inference engine, {\sc{MultiNest}} (\citealt{feroz_2008a} \& \citealt{feroz_2008b}), that has been optimized to sample efficiently from complex, degenerate, multi-peaked posterior distributions. {\sc{McAdam}} allows for the cluster and radio source/s (where present) parameters to be be fitted simultaneously directly to the short baseline (SB; $\sim 0.4-2$\,k$\lambda$) $uv$ data in the presence of receiver noise and primary CMB anisotropies. The high resolution, long baseline (LB; $\sim 2-10$\,k$\lambda$) data are used to constrain the flux and position of detected radio sources; these source-parameter estimates are then set as priors in the analysis of the SB data (see Section \ref{sec:rm}). Our short integration times required all of the LB data to be used for the determination of radio-source priors and none of the LB data were included in the {\sc{McAdam}} analysis of the SB data.
 Undertaking the analysis in the Fourier plane avoids the complications associated with going from the sampled visibility plane to the image plane. In {\sc{McAdam}}, predicted visibilities $V^{\rm{p}}_{\nu}(\mathbf{u}_i)$ at frequency $\nu$ and baseline vector ${\mathbf{u}}_i$, are generated and compared to the observed data through the likelihood function (see \citealt{feroz_2009} for a detailed overview). 

The observed SZ surface brightness towards the cluster electron reservoir can be expressed as
\begin{equation}
\Delta I_{\rm{CMB}}= \Delta T_{\rm{CMB}} \frac{dB(\nu,T)}{dT}\Bigg|_{T_{\rm{CMB}}} 
\end{equation}
where  $\frac{dB(\nu,T)}{dT}\Large|_{T_{\rm{CMB}}}$ is the derivative of the black body function at $T_{\rm{CMB}}$ -- the temperature of the CMB radiation (\citealt{fixsen1996}).
The CMB brightness temperature from the SZ effect is given by
\begin{equation}
\Delta T_{\rm{CMB}} = f(\nu)y T_{\rm{CMB}}.
\label{eq:deltaT}
\end{equation}
Here, $f(\nu)$ is the frequency ($\nu$)-dependent term of the SZ effect,
\begin{equation}
f(\nu) = \left( x\frac{e^x+1}{e^x-1} - 4 \right) (1 + \delta_{\rm{SZ}}(x,T_{\rm{e}})),
\end{equation}
where the $\delta_{\rm{SZ}}$ term accounts for relativistic corrections (see \citealt{itoh1998}), $T_{\rm{e}}$ is the electron temperature, $x=h\nu/{\rm{k}}_{\rm{B}}T_{\rm{CMB}}$, $h$ is Planck's constant and $\rm{k_B}$ is the Boltzmann constant.
To calculate the contribution of the cluster SZ signal to the (predicted) visibility data, the Comptonization parameter, $y$, across the sky must be computed:
\begin{equation}
y(s) = \frac{\sigma_{\rm{T}}}{m_{\rm{e}}c^2} \int^{+\infty}_{-\infty} n_{\rm{e}}(r)\rm{k_B}T_{\rm{e}}(r)\rm{d}l \propto \int^{+\infty}_{-\infty} P_{\rm{e}}(r)\rm{d}l .
\label{eq:littley}
\end{equation}
Here, $\sigma_{\rm{T}}$ is the Thomson scattering cross-section, $m_{\rm{e}}$ is the electron mass, $n_{\rm{e}}(r)$, $T_{\rm{e}}(r)$ and $P_{\rm{e}}(r)$ are the electron density, temperature and pressure at radius $r$ respectively, $c$ is the speed of light and $\rm{d}l$ is the line element along the line of sight. 
The projected distance from the cluster center to the sky is denoted by $s$, such that $r^2=s^2+l^2$.
The integral of $y$ over the solid angle $\rm{d}\Omega$ subtended by the cluster is proportional to the volume-integrated gas pressure, meaning this quantity correlates well with the mass of the cluster. For a spherical geometry this is given by
\begin{equation}
Y_{\rm{sph}}(r) = \frac{\sigma_{\rm{T}}}{m_{\rm{e}}c^2} \int^{r}_{0} P_{\rm{e}}(r')4\pi r'^2 \rm{d}r'.
\label{eq:inty}
\end{equation}
When $r \rightarrow \infty$, Equation \ref{eq:inty} can be solved analytically, as shown in \cite{perrott2014}, yielding the total integrated Compton-$y$ parameter, $Y_{\rm{T,phys}}$, which is related to the SZ surface brightness integrated over the cluster's extent on the sky through the angular diameter distance to the cluster ($D_{\rm{A}}$) as $Y_{\rm{T}}=Y_{\rm{T,phys}}/D_{\rm{A}}^2$.

\begin{table}
\centering
\caption{Summary of the cluster priors used in our analysis for the observational gNFW cluster parameterization (model I, Section \ref{sec:cm1}). 
$\Delta x_{\rm{c}}$ and $\Delta y_{\rm{c}}$ are the displacement from the map center to the centroid of the SZ decrement in RA and Dec, respectively. $\eta$ is the ellipticity parameter and $\Phi$ the position angle. $\theta_s=r_s/D_{\rm{A}}$, where $r_s$ is the scale radius and $D_{\rm{A}}$ is the angular size distance to the cluster. $Y_{\rm{T}}$ is the SZ surface brightness integrated over the cluster's extent on the sky. The $\theta_s$ and $Y_{\rm{T}}$ priors have been previously used in \protect\cite{ESZ} and \protect\cite{PlanckAMI}.}
\label{Tab:cluster_prior_gNFW}
\begin{tabular}{lcc}
\hline \hline
Parameter                                  & Prior                                        \\ \hline 
$\Delta x_{\rm{c}}$                               & Gaussian centered at pointing centre, $\sigma= 60\arcsec$\\
$\Delta y_{\rm{c}}$ 			      & Gaussian centered at pointing centre, $\sigma= 60\arcsec$\\
$\eta$ 			               & Uniform from 0.5 to 1.0 \\	
$\Phi$				      & Uniform from 0 to $180^{\circ}$\\	
$\theta_s$				      & $\lambda e^{-\lambda \theta_s}$  for\\
                                                     & $1.3' < \theta_s < 45'$ \& 0 outside this range\\
 $Y_{\rm{T}}$			      &	$Y_{\rm{T}}^{-\alpha}$ for $0.0005$ to $0.2 \rm{arcmin}^2$ \\
					      &  \& 0 outside this range with $\alpha = 1.6$\\      \hline
\end{tabular}
\end{table}

\subsection{Models and Parameter Estimates}
\label{sec:cp}

Analyses of X-ray or SZ data of the intra-cluster medium (ICM) that aim to estimate cluster parameters are usually based on a parameterized cluster model.  Cluster models necessarily assume a geometry for the SZ signal, typically spherical, and functional forms of two linearly-independent thermodynamic cluster quantities such as electron temperature and density. These models commonly make assumptions such as, the cluster gas is in hydrostatic equilibrium or that the temperature or gas fraction throughout the cluster is constant. Consequently, the accuracy and validity of the results will depend on how well the chosen parameterization fits the data and on the effects of the model assumptions (see e.g., \citealt{plagge2010}, \citealt{mroc2011} \& \citealt{rodriguez2011} for studies exploring model effects in analyses of real data and \citealt{olamaie2012} and \citealt{olamaie2013} for similar work on simulated data). 
In this work we present cluster parameters calculated from two different models; one is based on a fixed-profile-shape gNFW parameterization, for which typical marginalised parameter distributions for similar interferometric data from AMI have been shown in e.g., \cite{perrott2014}, and a second is based on the $\beta$ profile with variable shape parameters, where typical marginalised parameter distributions for comparable AMI data have been presented in \cite{carmen2012}. Comparison of marginalised posteriors for CARMA and AMI data in \cite{shimwell2013} for the $\beta$ model showed the distributions to be very similar. The clusters presented here are at modest redshifts and are unlikely to be in hydrostatic equilibrium - adopting two models at least allows the dependency of the cluster parameters on the adopted model to be illustrated and a
comparison with previous work to be undertaken.

\subsubsection{Cluster model I: observational gNFW parameterization}
\label{sec:cm1}

For cluster model I, we have used a generalized-NFW (gNFW; \citealt{navarro1996}) pressure profile in the same fashion as in the analysis of {\it{Planck}} data \citep{ESZ} to facilitate comparison of cluster parameters. 
A gNFW pressure profile with a fixed set of parameters is believed to be a reasonable choice since (1) numerical simulations show low scatter amongst cluster pressure profiles, with the pressure being one of the cluster parameters that suffers least from the effects of non-gravitational processes in the ICM out to the cluster outskirts and (2) the dark matter potential plays the dominant role in defining the distribution of the gas pressure, yielding a (pure) NFW form to the profile, which can be modified into a gNFW form to account for the effects of ICM processes (see e.g., \citealt{vikhlinin2005} and \citealt{nagai2007}). Using a fixed gNFW profile for cluster models has become regular practice (e.g.,  \cite{atrio2008} for WMAP, \cite{mroc2009} for SZA,  \cite{nicole} for BOLOCAM and \cite{plagge2010} for SPT data).

Assuming a spherical cluster geometry, the form of the gNFW pressure profile is the following:
\begin{equation}
P_{\rm{e}}(r) = P_{0}\left( \frac{r}{r_s}\right)^{-c} \left[ 1 + \left( \frac{r}{r_s} \right)^{a} \right]^{(c - b)/a},
\label{eq:gNFW}
\end{equation}
where $P_{0}$ is the normalization coefficient of the pressure profile and $r_s$ is the scale radius, typically expressed in terms of the concentration parameter $c_{500}=r_{500}/r_s$. Parameters with a numerical subscript 500, like $c_{500}$, refer to the value of that variable within $r_{500}$---the radius at which the mean density is 500 times the critical density at the cluster redshift. 
The shape of the profile at intermediate regions ($r \approx r_s$), around the cluster outskirts ($r >> r_s$) and in the core regions ($r << r_s$) is governed by three parameters $a$, $b$, $c$, respectively. Together with $c_{500}$, they constitute the set of gNFW parameters. Two main sets of gNFW parameters have been derived from studies of X-ray observations (inner cluster regions) and simulations (cluster outskirts) (\citealt{nagai2007} and \citealt{arnaud2009}). For ease of comparison with the ${\it{Planck}}$ results, as well as with SZ-interferometer data e.g., from AMI in \cite{PlanckAMI}, we have chosen to use the gNFW parameters derived by Arnaud et al.: $(c_{500}, a, b, c) = (1.156, 1.0620, 5.4807, 0.3292)$.

\begin{table*}
\setlength{\tabcolsep}{2mm}
\centering
\small
\caption{Mean and 68\%-confidence uncertainties for {\sc{McAdam}}-derived cluster parameters when fitting for an observational gNFW cluster parameterization (Model I; Section \ref{sec:cm1}) for clusters with a significant SZ detection in the CARMA-8 data (Table \ref{tab:evidences}). The cluster ID is a shorthand naming convention adopted here and in Paper 1, since not all our targets have an identifier in the {\sc{Planck}} Union catalog (\citealt{Planck2013}). Where available, the Union catalog names are given in Table \ref{tab:clusters}. The derived sampling parameters for the gNFW parameterization are presented in columns 2 to 7 and their priors are listed in Table \ref{Tab:cluster_prior_gNFW}. $Y_{500}$ is the integrated SZ surface brightness within $\theta_{500}$,  where $\theta_{500}=r_{500}/D_{\rm{A}}$ and $y(0)$ is the central Comptonization parameter, $y$, Equation \ref{eq:littley}. }
\label{tab:paramsGNFW}
{\renewcommand{\arraystretch}{1.8}
\begin{tabular}{lcccccccccc}
\hline \hline
 Cluster ID & $\Delta x_{\rm{c}}$ & $\Delta y_{\rm{c}}$ & $\Phi$ & $\eta$ & $\theta_s$ & $Y_{\rm{T}}$ & $\theta_{500}$ & $Y_{500}$      &   $ y(0)$ \\
   & \arcsec & \arcsec & deg &  & \arcmin & $\times 10^{-4}\,\,\,\rm{arcmin}^2$ & \rm{arcmin} &$\times 10^{-4}\,\,\, \rm{arcmin}^2$ & $\times 10^{-4}$\\ \hline
P014 & $25^{+24}_{-23}$ & $-148^{+6}_{-14}$ & $149^{+22}_{-22}$ & $0.6^{+0.1}_{-0.1}$ & $4^{+1}_{-1}$ & $20^{+10}_{-11}$ & $4.3^{+1.2}_{-1.3}$ & $11^{+5}_{-6}$ & $1.6^{+0.4}_{-0.4}$\\
P086& $68^{+20}_{-20}$ & $91^{+17}_{-17}$ & $85^{+71}_{-65}$& $0.8^{+0.2}_{-0.2}$ & $3^{+1}_{-2}$ & $20^{+6}_{-15}$ & $3.7^{+1.6}_{-2.2}$ & $11^{+3}_{-8}$ & $2.1^{+0.9}_{-0.8}$\\
P097&$79^{+23}_{-23}$ & $38^{+16}_{-17}$ & $76^{+66}_{-52}$ & $0.8^{+0.2}_{-0.2}$ & $3^{+1}_{-1}$ & $11^{+3}_{-6}$ & $3.2^{+1.2}_{-1.3}$ & $6^{+2}_{-4}$ & $1.9^{+0.9}_{-0.8}$\\
P109&$10^{+18}_{-18}$ & $75^{+14}_{-13}$ & $89^{+91}_{-89}$ & $0.8^{+0.2}_{-0.2}$ & $3^{+1}_{-1}$ & $10^{+3}_{-5}$ & $3.1^{+1}_{-1.1}$ & $6^{+2}_{-3}$ & $1.9^{+0.8}_{-0.8}$\\
P170&$-59^{+12}_{-12}$ & $10^{+14}_{-14}$ & $69^{+27}_{-26}$ & $0.7^{+0.2}_{-0.2}$ & $2^{+1}_{-1}$ & $12^{+3}_{-7}$ & $2.7^{+0.9}_{-1.2}$ & $6^{+2}_{-4}$ & $2.3^{+0.8}_{-0.7}$\\
P187&$64^{+14}_{-14}$ & $-67^{+15}_{-15}$ & $101^{+50}_{-59}$ & $0.8^{+0.2}_{-0.2}$ & $4^{+2}_{-2}$ & $23^{+10}_{-18}$ & $4.1^{+1.8}_{-1.9}$ & $13^{+5}_{-10}$ & $1.9^{+0.6}_{-0.6}$\\
P190&$59^{+10}_{-10}$ & $11^{+11}_{-11}$ & $95^{+40}_{-39}$ & $0.8^{+0.2}_{-0.1}$ & $3^{+1}_{-1}$ & $16^{+7}_{-11}$ & $3.5^{+1.4}_{-1.4}$ & $9^{+4}_{-6}$ & $2.0^{+0.6}_{-0.6}$\\
P205&$-83^{+11}_{-11}$ & $-26^{+15}_{-15}$ & $79^{+23}_{-23}$ & $0.7^{+0.2}_{-0.1}$ & $4^{+2}_{-2}$ & $31^{+13}_{-26}$ & $4.9^{+2.1}_{-2.2}$ & $17^{+7}_{-15}$ & $1.6^{+0.4}_{-0.4}$\\
P351&$-42^{+34}_{-34}$ & $63^{+24}_{-23}$ & $71^{+60}_{-45}$ & $0.7^{+0.2}_{-0.2}$ & $5^{+2}_{-2}$ & $19^{+7}_{-14}$ & $5.3^{+2.3}_{-2.3}$ & $10^{+4}_{-8}$ & $0.9^{+0.3}_{-0.4}$\\
\hline
\end{tabular}
}
\end{table*}

In our gNFW analysis, we characterize the cluster by the following set of sampling parameters (Table \ref{Tab:cluster_prior_gNFW}):
\begin{equation*}
 {\bm{\mathcal{P}}}_{\rm{c}} = (\Delta x_{\rm{c}}, \Delta y_{\rm{c}}, \eta, \Phi, \theta_s = r_s/D_{\rm{A}}, Y_{\rm{T}}).
 \end{equation*} 
Here, $\Delta x_{\rm{c}}, \Delta y_{\rm{c}}$ are the displacement of the cluster decrement from the pointing centre, where the cluster right ascension is equal to the map center (provided in Table \ref{tab:clusters}),  $\eta$ is the ellipticity parameter, that is, the ratio of the semi-minor and semi-major axes and $\Phi$ is the position angle of the semi-major axis, measured N through E i.e. anti-clockwise. We note that the projected cluster decrement is modeled as an ellipse and hence our model is not properly triaxial. 

The priors used in this analysis are given in Table \ref{Tab:cluster_prior_gNFW}; they have been used previously for the blind detection of clusters in {\it{Planck}} data \citep{ESZ} and to characterize confirmed and candidate clusters in \cite{PlanckAMI}. Cluster parameter estimates and the CARMA best-fit positions derived from model I are provided in Tables \ref{tab:paramsGNFW} and \ref{tab:newpos}, respectively.

\begin{table}
\centering
\small
\caption{Cluster J2000 coordinates derived using the gNFW (model I) fits to the CARMA-8 data.}
\label{tab:newpos}
\begin{tabular}{lccc}
\hline \hline
Cluster ID & RA  & Dec \\ 
           & hh:mm:ss & dd:mm:ss \\ \hline          
P014 & 16:03:23.29  &  03:16:44.00 \\
P086 & 15:14:00.85  &   52:48:12.56  \\       
P097 & 14:55:24.17  &  58:52:20.44\\
P109 & 18:23:03.50  &  78:23:07.19\\
P170 & 08:50:59.16   & 48:30:28.14  \\     
P187 & 07:32:23.03   &  31:37:32.03\\
P190 & 11:06:08.81   & 33:33:56.23\\
P205 & 11:38:07.21   & 27:54:39.62 \\
P351 & 15:04:02.09   & -06:06:12.24\\
\hline
\end{tabular}
\end{table}

 \subsubsection{Cluster model II: observational $\beta$ parameterization}
\label{sec:cm2}

For this cluster parameterization we fit for an elliptical cluster geometry, as we did for model I, and model the shape of the SZ temperature decrement with a $\beta$-like profile \citep{Cavaliere1978}:
\begin{equation}
\Delta T_{\rm{CMB}}(\theta) = \Delta T_0 \left( 1 + \left(\frac{\theta}{\theta_{\rm{c}}} \right)^2 \right)^{\frac{1-3\beta}{2}}, 
\label{eq:nebeta}
\end{equation}
where $\Delta T_0$ is the brightness temperature decrement at zero projected radius, while $\beta$ and $r_{\rm{c}}=\theta_{\rm{c}}\times D_{\rm{A}}$---the power law index and the core radius---are the shape parameters that give the density profile a flat top at small $\frac{\theta}{\theta_{\rm{c}}}$ and a logarithmic slope of $3\beta$ at large $\frac{\theta}{\theta_{\rm{c}}}$. 
The sampling parameters for the cluster signal are:
\begin{equation*}
 {\bm{\mathcal{P}}}_{\rm{c}}=(\Delta x_{\rm{c}}, \Delta y_{\rm{c}}, \eta, \phi,\Delta T_0, \beta, \theta_{\rm{c}}),
\end{equation*}
with priors given in Table \ref{tab:cluster_prior_BLOB}, which allow for the $y$ signal to be computed. Cluster parameter estimates derived from model II are provided in Table \ref{tab:paramsBLOB}.

\begin{table}
\centering
\caption{Summary of the cluster priors used in our analysis for the observational $\beta$ cluster parameterization (model II, Section \ref{sec:cm2}). $\Delta x_{\rm{c}}$ and $\Delta y_{\rm{c}}$ are the displacement from the map center to the centroid of the SZ decrement in RA and Dec, respectively. $\eta$ is the ellipticity parameter and $\Phi$ the position angle. The power law index $\beta$ and the core radius $r_{\rm{c}}$ are the shape parameters of the density profile and $\Delta T_0$ is the temperature decrement at zero projected radius.}
\label{tab:cluster_prior_BLOB}
\begin{tabular}{lcc}
\hline \hline
Parameter                                  & Prior                                        \\ \hline 
$\Delta x_{\rm{c}}$                               & Gaussian centered at pointing centre, $\sigma= 60''$\\
$\Delta y_{\rm{c}}$ 			      & Gaussian centered at pointing centre, $\sigma= 60''$\\
$\eta$ 			               & Uniform from 0.5 to 1.0 \\	
$\Phi$				      & Uniform from 0 to $180^{\circ}$\\
$\beta$                                       & Uniform  from  $0.4$ to $2.5$                    \\
$\theta_{\rm{c}}$                       & Uniform  from  $20$ to $500\arcsec$                 \\
$\Delta T_0$                             & Uniform  from   $-3$ to $-0.01$\,mK              \\ \hline
\end{tabular}
\end{table}

\begin{table*}
\centering
\small
\caption{Mean and 68\%-confidence uncertainties for {\sc{McAdam}}-derived cluster parameters when fitting for an observational $\beta$ cluster parameterization  (Model II; see Section \ref{sec:cm2}) for clusters with an SZ detection in the CARMA-8 data (see Table \ref{tab:evidences}). The priors for these sampling parameters are given in Table \ref{tab:cluster_prior_BLOB}.}
{\renewcommand{\arraystretch}{1.8}
\label{tab:paramsBLOB}
\begin{tabular}{lcccccccc}
\hline \hline
Cluster ID & $\Delta x_{\rm{c}}$ & $\Delta y_{\rm{c}}$ & $\Phi$  & $\eta$ & $\theta_{\rm{c}}$ & $\beta$ & $\Delta T_0$   \\ 
 		& \arcsec 			& \arcsec			  & deg      &		& \arcsec			&		& microK \\ \hline
 P014   & $52^{+6}_{-6}$  & $-137^{+7}_{-9}$   & $90^{+10}_{-10}$ &$0.7^{+0.3}_{-0.2}$    &$91^{+14}_{-71}$ & $1.6^{+0.9}_{-0.9}$   & $-682^{+145}_{-67}$\\
 P086   & $76^{+15}_{-15}$  & $83^{+15}_{-14}$ & $102^{+78}_{-102}$ &$0.8^{+0.2}_{-0.3}$    &$82^{+10}_{-62}$   & $1.7^{+0.8}_{-1.0}$ & $-1229^{+415}_{-153}$\\
 P097  & $77^{+8}_{-11}$    & $34^{+6}_{-6}$   & $82^{+98}_{-82}$ &$0.8^{+0.2}_{-0.3}$     &$78^{+4}_{-58}$ & $1.8^{+0.7}_{-1.1}$ & $-1109^{+505}_{-218}$\\
 P109    & $7^{+5}_{-5}$    & $60^{+5}_{-4}$   & $85^{+21}_{-85}$ &$0.8^{+0.2}_{-0.3}$     &$60^{+4}_{-40}$ & $1.9^{+0.6}_{-0.1}$ & $-1435^{+357}_{-259}$\\
 P170    & $-58^{+5}_{-6}$     & $4^{+7}_{-7}$    & $63^{+4}_{-11}$ &$0.7^{+0.3}_{-0.2}$   &$112^{+23}_{-92}$   & $1.5^{+1.0}_{-0.8}$    & $-816^{+231}_{-8}$\\
 P187   & $52^{+6}_{-6}$   & $-48^{+7}_{-6}$   & $92^{+88}_{-92}$ &$0.8^{+0.2}_{-0.3}$   &$111^{+20}_{-91}$   & $1.7^{+0.8}_{-1.0}$    & $-706^{+172}_{-7}$\\
 P190   & $53^{+4}_{-4}$    & $21^{+5}_{-5}$  & $74^{+106}_{-74}$ &$0.8^{+0.2}_{-0.3}$    &$78^{+13}_{-58}$   & $1.5^{+1.0}_{-0.8}$   & $-959^{+311}_{-61}$\\
 P205    & $-80^{+6}_{-6}$ & $-19^{+10}_{-11}$     & $74^{+4}_{-7}$   &$0.6^{+0.1}_{-0.1}$  &$205^{+39}_{-185}$ & $1.3^{+1.2}_{-0.6}$   & $-846^{+304}_{-11}$\\
 P351  & $15^{+20}_{-21}$  & $35^{+18}_{-15}$    & $55^{+10}_{-10}$   &$0.6^{+0.1}_{-0.1}$ &$291^{+209}_{-271}$   & $1.5^{+1.0}_{-0.8}$  & $-959^{+439}_{-175}$\\
\hline
\end{tabular}
}
\end{table*}

\begin{figure*}
\begin{center}
\centerline{\includegraphics[width=8.0cm, height=8.0cm,clip=,angle=0.]{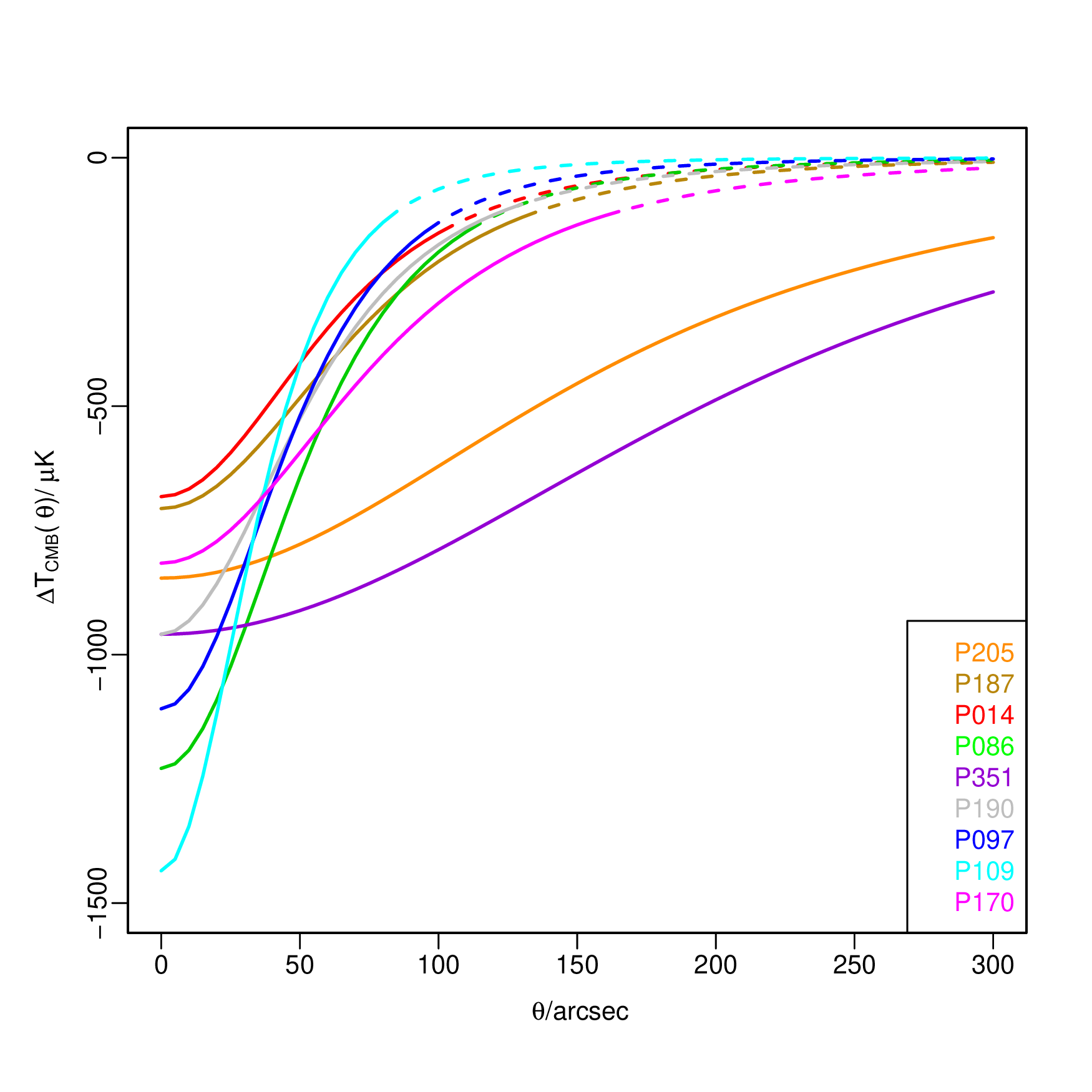}\qquad\includegraphics[width=8.0cm,height=8.0cm,clip=,angle=0.]{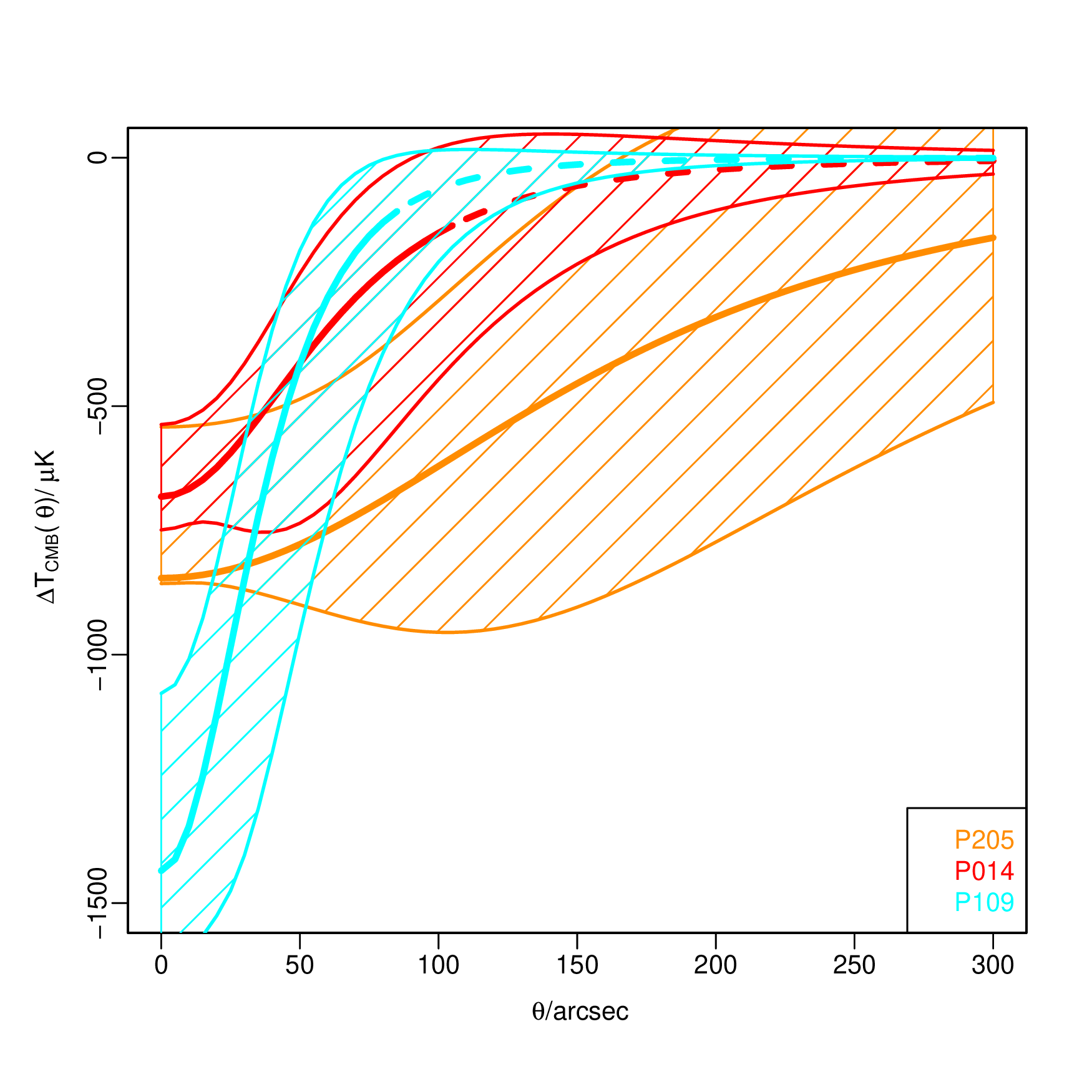}}

\caption{Left: radial brightness temperature profiles derived from Equation \ref{eq:nebeta} using the cluster parameter values ($\Delta T_0$, $\beta$ and $\theta_{\rm{c}}$) from fits of model II to the CARMA-8 data (Table \ref{tab:paramsBLOB}). When the profile has dropped to $3$ times the noise in the SB data, the lines turn from solid to dashed. In the legend, the cluster candidates have been ordered by decreasing $Y_{500}$ (Table \ref{tab:paramsGNFW}), although it should be noted that for some clusters the difference in $Y_{500}$ is small. Shallower profiles with the most negative $\Delta T_0$ values and largest $\theta_{500}$ values should correspond to the largest $Y_{500}$ from model I. Comparing the volume-integrated brightness temperature profile, $\int T(\theta) 4\pi\theta^2 \rm{d}\theta$ for each cluster from 0 to its $\theta_{500}$ (from model I; Table \ref{tab:paramsGNFW}) with $Y_{500}$ (model I), shows reasonable correspondence for most systems, with the exception of P351 and P187.
 Right: best-fit radial brightness temperature profiles for three clusters, P205, P014 and P109, which span a range of profile shapes (in thick solid lines, same as the corresponding lines in the
 left panel). However, here we include the upper and lower $68\%$-confidence limits of each brightness temperature profile in thin solid lines and highlight the region between these limits with diagonal lines. This shows that, according to the parameter fits to CARMA-8 data from model II, the clusters in our sample exhibit a heterogenous set of profiles, distinguishable despite the uncertainty. Furthermore, high resolution SZ data or X-ray data may be insensitive to the large signal from the outer parts of the cluster and introduce an additional uncertainty in the derivation
 of $Y_{T}$, an issue which is addressed through a joint analysis with the {\it Planck} data.}
\label{fig:tprofile}
\end{center}
\end{figure*}

It is important to note that, historically, in many SZ analyses the shape of the $\beta$ profile has been fixed to values obtained from fits to higher resolution X-ray data. However, these X-ray results primarily probe the inner regions of the cluster and, thus, can provide inadequate best-fit profile shape parameters for SZ data extending out to larger-$r$. A comparative analysis in \cite{czakon2014} reveals systematic differences in cluster parameters derived from SZ data using a model-independent method versus X-ray-determined cluster profiles.
Several studies have now shown that fits to SZ data reaching $r_{500}$ and beyond preferentially yield larger $\beta$ values than X-ray data, which tend to yield $\beta = 2/3$ (see e.g., \citealt{tash2012} \& \citealt{plagge2010}). Using a suitable $\beta$ (and $\theta_{\rm{c}}$) value for the aforementioned typical SZ data can yield results comparable to those of a gNFW profile.  In our $\beta$ parameterization we allow the shape parameters, $\beta$ and $\theta_{\rm{c}}$, to be fit in {\sc{McAdam}}, since they jointly govern the profile shape. While our data cannot constrain either of these variables independently, they can constrain their degeneracy.

Although a large fraction of clusters are well-described by the best-fit gNFW parameterizations, some are not, as can be seen from e.g., the spread in the gNFW parameter sets from fits to individual clusters in the REXCESS sample \citep{arnaud2009}. In these cases, modelling the cluster using a fixed (inadequate) set of gNFW parameters will return biased, incorrect  results, whereas using a $\beta$ model with varying $\beta$ and $\theta_{\rm{c}}$ should provide more reliable results. 
This is shown in Figure 1 of \cite{carmen2012} where data from AMI for a relaxed and a disturbed cluster are analyzed with a $\beta$ parameterization and five gNFW parameterizations, four of which have gNFW sets of parameters drawn from the Arnaud et al. REXCESS sample, three from individual systems and one from the averaged (Universal) profile, and, lastly, one with the average-profile values from an independent study by \cite{nagai2007}. For both clusters, the Nagai parameterization lead to a larger $Y_{500}-\theta_{500}$ degeneracy and larger parameter uncertainties than the Arnaud Universal parameterization. The mean $Y_{500}$ and $\theta_{500}$ values obtained from using the $\beta$, Universal (Arnaud) and Nagai gNFW profiles were consistent to within the 95\% probability contours, but this was not the case for fits using the sets of gNFW parameters obtained from individual fits to REXCESS clusters, indicating that some clusters do not follow a single, averaged profile. Here, comparison of the Bayesian evidence values for beta and gNFW-based analyses showed that the data could not distinguish between them. Our CARMA data for this paper have a similar resolution to the AMI data but, typically, they have much poorer SNRs and similarly cannot determine which of the two profiles provides a better fit to the data. More recently, \citet{sayers2013b} have derived a new set of gNFW parameters from 45 massive galaxy clusters using Bolocam and \citet{mantz2014} have further shown how the choice of model parameters can have a measurable effect on the estimated Y-parameter. 

All sets of gNFW parameters can lead to biases when applied to different sets of data. Given that there is no optimally-selected set of gNFW parameters to represent CARMA 31-GHz data towards massive, medium-to-high redshift clusters (z $\gtrsim$0.5), we choose to base most of our analysis on the gNFW parameter set from the 'Universal' profile derived by \citep{arnaud2009} as this facilitates comparison with the Planck analysis and parallel studies between Planck and AMI, an interferometer operating at 16 GHz with arcminute resolution.

 \paragraph{Cluster Profiles}
\label{sec:clusprof}

Using Equation \ref{eq:nebeta} for $\Delta T_{\rm{CMB}}(\theta)$ (the SZ temperature decrement), and the mean values for $\Delta T_0$, $\beta$ and $\theta_{\rm{c}}$ derived from model II fits to the CARMA-8 data (Table \ref{tab:paramsBLOB}), in Figure  \ref{fig:tprofile}, left, we plot the radial brightness temperature profiles for our sample of CARMA-8-detected candidate clusters. We order them in the legend by decreasing CARMA-8 $Y_{500}$, from Table \ref{tab:paramsGNFW}, although in some cases the differences are small.  We would expect clusters with the most negative $\Delta T_0$ values, the shallowest profiles and the largest $\theta_{500}$ to yield the largest $Y_{500}$ values. While there is reasonable correspondence throughout our cluster sample, two clusters P351 and P187 are outliers in this relation. Computing $\int T(\theta) 4\pi\theta^2 \rm{d}\theta$ for each cluster from 0 to its $\theta_{500}$, determined from model I (Table \ref{tab:paramsGNFW}) shows that P351 (P187) has the highest (fifth highest) volume-integrated brightness temperature profile but only the fifth highest (second highest) $Y_{500}$ .  

In Figure  \ref{fig:tprofile}, right, we plot the upper and lower limits of the brightness temperature profiles allowed by the profile uncertainties for three clusters, P014, P109  and P205, chosen to span a wide range of profile shapes. It can be seen that the cluster candidates display a range of brightness temperature profiles that can be differentiated despite the uncertainties. In Table \ref{tab:profratio}, by computing the ratio of the integral of the brightness temperature profile within (a) the 100-GHz {\it{Planck}} beam and (b) $\theta_{500}$ from Table \ref{tab:paramsGNFW}, we quantify how concentrated each brightness temperature profile is. The profile concentration factors have a spread of a factor of $\approx 2.5$ but for six clusters they agree within a factor of $\approx 1.2$.  Furthermore, the derived ellipticities shown in Table \ref{tab:paramsBLOB} and \ref{tab:paramsGNFW}, which can be constrained by the angular resolution of the CARMA-8 data, show significant evidence of morphological irregularity suggesting that these clusters may be disturbed and heterogeneous systems.

\subsubsection{Radio-source Model and Parameter Estimates}
\label{sec:rm}

Radio sources are often strong contaminants of the SZ decrement and their contributions must be included in our cluster analysis. In this work, we jointly fit for the cluster, radio source and primary CMB signals in the SB data. The treatment of radio sources is the same for all cluster models. These sources are parameterized by four parameters,
\begin{equation*}
\bm{\Theta_{\rm{s}}} = ({\rm{RA}}_{\rm{s}}, {\rm{Dec}}_{\rm{s}}, \alpha_{\rm{s}}, S_{31}),
\end{equation*}
where $\rm{RA}_{\rm{s}}$ and $\rm{Dec}_{\rm{s}}$ are RA and Dec of the radio source, $\alpha_{\rm{s}}$ is the spectral index, derived from the low fractional CARMA-8 bandwidth and $S_{31}$ is the 31-GHz integrated source flux. We adopt the $S_{\nu} \propto \nu^{-\alpha_{\rm{s}}}$ convention, where $S$ is flux and $\nu$ frequency. 

The high resolution LB data were mapped in {\sc{Difmap}} (\citealt{shepherd1997}) to check for the presence of radio sources. Radio-point sources detected in the LB maps were modeled using the {\sc{Difmap}} task {\sc{Modelfit}}. 
The results from {\sc{Modelfit}} were primary-beam corrected using a FWHM of 660 arcseconds by dividing them by the following factor:
\begin{equation}
\rm{exp}\left( \frac{-r^2}{2\times\sigma^2}\right),
\end{equation}
where $r$ is the distance of the source to the pointing centre and 
\begin{equation}
\sigma= \frac{660}{2 \times (2 \times \ln(2))^{0.5}} .
\end{equation}
The primary-beam corrected values were used as  priors in the analysis of the SB data (see Table \ref{tab:source_priors}). {\sc{Modelfit}} values are given in Paper 1 and McADam-derived values are provided in Table \ref{tab:sparams}.

\begin{table}
\centering
\small
\caption{Ratio of the integral of $\int T(\theta)d\theta$ integrated between 0 and $600\arcsec$ ($\approx$FWHM of the 100-GHz {\it{Planck}} beam) and integrated between 0 and $\theta_{500}$ (from Table \ref{tab:paramsGNFW}), where the expression for $T(\theta)$ is given in Equation \ref{eq:nebeta}. This ratio is a measure of concentration of the brightness temperature profile. The most concentrated profile (for P014) is $\approx 2.5$ times more concentrated than the shallowest profile (for P351). Six of the nine clusters are equally concentrated to within a factor of $\approx 1.2$.}
\label{tab:profratio}
\begin{tabular}{lcc}
\hline \hline
 Cluster ID & $\frac{\int^{600\arcsec}_0 T(\theta) d\theta}{\int^{\theta_{500}}_0 T(\theta) d\theta}$ \\ \hline 
   P014 & 17.3\\
   P086 & 17.1\\
   P097& 17.8\\ 
   P109& 13.5\\ 
   P170& 36.0\\ 
   P187& 20.8\\ 
   P190 & 19.4\\
   P205 & 40.5\\
   P351& 45.07 \\ \hline
\end{tabular}
\end{table}

\begin{table}
\caption{Summary of the source priors used in our analysis. Values for the position ($x_{\rm{s}}$, $y_{\rm{s}}$) and 31-GHz flux ($S_{31}$) priors were obtained from the long baseline CARMA-8 data (see Paper 1 and Section \ref{sec:rm} for further details). The error on the source location is the typical error of the LB data. To account for the fact that in the combined
cluster and source analysis uses only the lower-resolution data, we model the integrated 31-GHz source flux with a Gaussian. The width of this Gaussian is set to $\sigma= 20\%$ of the source flux. For the spectral index $\alpha_{\rm{s}}$ , we used a wide prior, encompassing reasonable values (see Section \ref{sec:rsc}). }
\label{tab:source_priors}
\begin{tabular}{lcc}
\hline \hline
Parameter                                   & Prior \\ \hline
${\rm{RA}}_{\rm{s}}$, ${\rm{Dec}}_{\rm{s}}$        & Uniform between $\pm10\arcsec$ \\
                                                       & from the LB-determined position\\
$S_{31}$                                 & Gaussian centered at best-fit {\sc{Modelfit}} value\\
                                                      &  with a $\sigma$ of 20$\%$ \\
$\alpha_{\rm{s}}$                        & Gaussian centered at 0.6 with $\sigma = 0.5$ \\  \hline
\end{tabular}
\end{table} 

\subsection{Quantifying the Significance of the CARMA-8 SZ Detection or Lack thereof}
\label{sec:detectability}

Bayesian inference provides a quantitative way of ranking model fits to a dataset. Although the term model technically refers to a position in parameter space $\bm{\Theta}$, here
we refer to two model $\it{classes}$: a model class that allows for a cluster signal to be fit to the data, $M_{1}$, and another, $M_0$, that does not.
The parameterization we have used for this analysis has been the gNFW-based model, Model I;  for the $M_1$ case Model I was run as described in Section \ref{sec:cm1} and for the $M_0$ case it was run in the same fashion except for the prior on $Y_{500}$, which was set to 0, such that no SZ (cluster) signal is included in the model. Given the data ${\bf{D}}$, deciding whether $M_1$ or $M_0$ fit the data best can be done by computing the ratio:

\begin{equation}
 \frac{\rm{Pr}(M_1|\mathbf{D})}{\rm{Pr}(M_0|\mathbf{D})}= \frac{\rm{Pr}(\mathbf{D}|M_1)}{\rm{Pr}(\mathbf{D}|M_0)} \frac{\rm{Pr}(M_1)}{\rm{Pr}(M_0)} 
= \frac{Z_1}{Z_0} \frac{\rm{Pr}(M_1)}{\rm{Pr}(M_0)}.
\end{equation}
Here, $K=\frac{\rm{Pr}(\mathbf{D}|M_1)}{\rm{Pr}(\mathbf{D}|M_0)}$ is known as the Bayes Factor and $\frac{\rm{Pr}(M_1)}{\rm{Pr}(M_0)}$ is the prior ratio, that is, the probability ratio of the two model classes, which must be set before any information has been drawn from the data being analyzed. Here, we set the prior ratio to unity\footnote{Prior ratios need not be set to unity see e.g, \cite{jenkins2011}.} i.e. we assume no {\it{a priori}} knowledge regarding which model class is most favorable.  
The Bayesian Evidence $Z$ is calculated as the integral of the likelihood function, $\mathcal{L}= {\rm{Pr}}({\mathbf{D}}|{\bm{\Theta}},M)$, times the prior probability distribution $\Pi({\bm{\Theta}})={\rm{Pr}}({\bm{\Theta}}|M)$,
\begin{equation}
\begin{split}
Z & ={\rm{Pr}}({\mathbf{D}}|M)= \int {\rm{Pr}}({\mathbf{D}}|{\bm{\Theta}},M){\rm{Pr}}({\bm{\Theta}}|M)d^{\mathcal{D}}{\bm{\Theta}} \\
    & = \int {\bm{\mathcal{L}}}({\bm{\Theta}})\Pi({\bm{\Theta}}) {\rm{d}}^{\mathcal{D}}{\bm{\Theta}}
\end{split}
\end{equation}
where $\mathcal{D}$ is the dimensionality of the parameter space. $Z$ represents an average of the likelihood over the prior and will therefore favour models with high likelihood values throughout the entirety of parameter space. This satisfies OccamÕs razor, which states that the models with compact parameter spaces will have larger evidence values than more complex models, 
unless the latter fit the data significantly better i.e., unnecessary complexity in a model will be penalized with a lower evidence value. 

The derived Bayes factor is listed in Table \ref{tab:evidences} along with the corresponding classification of whether or not the cluster was considered to be detected. We find that all the SZ decrements considered to have high signal-to-noise ratios in Paper I have Bayes Factors that indicate the presence of a cluster signature is strongly favoured. However, we do find some tension between the Paper I {\sc{Modelfit}} and McAdam results for one of the candidate clusters, P014. In paper I this candidate cluster was catalogued as tentative (see Appendix B of Paper I for more details). The low SNR\footnote{The SNR for the CARMA SZ detections was calculated in Paper I as the ratio of the peak decrement, after correcting for beam attenuation, and the RMS of the SB data.} of 4.2 for the decrement together with the unusually large displacement from the {\it{Planck}} position ($\approx 159\arcsec$) suggest this detection is spurious. The lack of an X-ray signature would support this, unless it was a high-redshift cluster or one without a concentrated profile. With regards to the source environment, two sources were detected in the LB data with a peak 31-GHz flux density of 6.3 and 9.4 mJy, a distance of $\approx 100\arcsec$ and $\approx 600\arcsec$ from the SZ decrement, respectively. The LB data, after subtraction of these radio sources using the {\sc{Modelfit}} values, were consistent with noise-like fluctuations, indicating the removal of the radio-source flux worked well. The NVSS results revealed four other radio sources which, due to their location and measured fluxes at 1.4 GHz (as well as their lack of detection in the LB data), are unlikely to contaminate the candidate cluster. The NVSS results also indicate the radio sources are not extended. The strongest support for the presence of an SZ signature comes from the relatively-high {\it{Planck}} SNR of 4.5 but this measurement could suffer from the high contamination from interstellar medium emission which mimics itself as the SZ increment at high frequencies and could also result in a large error on its derived position -- suggestion of this arises from the strength of the 100$\mu$m emission which is the highest for our sample. Yet, despite these results, the Bayes factor from Table  \ref{tab:evidences} shows that a model with a cluster signature is preferred over one without. There are potentially quite important differences between the Paper I results and the McAdam evidences e.g., the {\sc{Modelfit}} results are based on an image and single-value fits without simultaneous fits to other parameters, while the McAdam results are derived from fits to the $uv$-plane, taking into account all model parameters, some of which are not strongly constrained by the data. Indeed, we find high scatter in the relation between evidence values and {\sc{Modelfit}}-based SNR values but they are positively correlated. Validation of this candidate cluster will require further data. Given the modest significance of the detection of P014 by two different techniques, we decided to include P014 in our {\sc{McAdam}} analyses.

\section{Constraints from {\it{Planck}}}
\label{sec:planck}

\subsection{Cluster Parameters}

We used the public {\it{Planck}} PR1 all-sky maps to derive $Y_{500}$ and $\theta_{500}$ values for our cluster candidates (Table \ref{tab:paramsY500}).  The values
were derived using a multifrequency matched filter \citep{melin2006, melin2012}. The $y$ profile from Equation \ref{eq:littley} is integrated over the cluster profile and then convolved with
the {\it Planck} beam at the corresponding frequency; the matched filter leverages only the {\it Planck} high frequency instrument (HFI) data between 100-857 GHz because it has been seen that the large beams
at lower frequencies result in dilution of the temperature decrement due to the cluster. The beam-integrated, frequency-dependent SZ signal is then fit with the scaled matched filter profile
from Equation \ref{eq:deltaT} to derive $Y_{500}$. The uncertainty in the derived $Y_{500}$ is due to both the uncertainty in the cluster size \citep{Planck2013}, as well as the signal to noise
of the temperature decrement in the {\it Planck} data. The large beam of {\it Planck}, FWHM$\approx 10\arcmin$ at 100\,GHz, makes it challenging to constrain the cluster size unless the clusters
are at low redshift and thereby significantly extended. For this reason, \citet{Planck2013} provided the full range of $Y_{500}-\theta_{500}$ contours which are consistent with the {\it Planck}
data. 

For the comparison here, there are two {\it{Planck}}-derived $Y_{500}$ estimates; $Y_{500}$ was calculated using the cluster position and size ($\theta_{500}$) obtained from the higher resolution CARMA-8 data, while $Y_{500,{\rm{blind}}}$ was computed using the {\it{Planck}} data {\it{alone}} without using the CARMA-8 size constraints. Similarly, $\theta_{500,{\rm{blind}}}$ is a measure of the angular size of the cluster using exclusively the {\it{Planck}} data; this value is weakly constrained and, thus, no cluster-specific errors are given for this parameter in Table \ref{tab:paramsY500}. At this point, it is important to note that the quoted 
uncertainty for $Y_{500, {\rm{blind}}}$ is an underestimate; the quoted error for this parameter is based on the spread in $Y_{500}$ at the best-fit $\theta_{500,{\rm{blind}}}$ and is proportional to the
signal to noise of the cluster in the {\it {Planck}} data i.e, without considering the error on $\theta_{500, blind}$ which is very large. However, the uncertainty in $Y_{500}$ is 
accurate since it propagates the true uncertainty in $\theta_{500}$ from the CARMA-8 data into the estimation of this quantity from the {\it {Planck}} maps.

The uncertainty in $Y_{500,{\rm{blind}}}$ from using the CARMA-8 size measurement has gone down by $\approx 60\%$ on average, despite the fact that the $Y_{500, {\rm{blind}}}$ does not include the uncertainty resulting from the unknown cluster size. If the true uncertainty in $Y_{500,{\rm{blind}}}$ had been taken into account, the uncertainty would have gone down by more than an order of magnitude after application
of the CARMA-8 derived cluster size constraints.
The mean ratio of $Y_{500, {\rm{blind}}}$ to $Y_{500}$ is 1.3 and, in fact, $Y_{500}$ is only larger than its blind counterpart for two systems. Differences in the profile shapes account for $Y_{500, {\rm{blind}}}$ being larger than $Y_{500}$ for three systems, P014, P097 and P187, for which $\theta_{500,{\rm{blind}}}$ is smaller than $\theta_{500}$ measured by CARMA-8.

\begin{table*}
\centering
\small
\caption{$Y_{500}$ and $\theta_{500}$ values derived from the release 1 {\it{Planck}} all-sky maps . The second and third columns contain $Y_{500}$  and $\theta_{500}$ from a blind analysis of the {\it{Planck}} maps, that is, from an analysis of {\it{Planck}} data {\it{alone}}, without any constraints from ancillary data. The fourth column contains the {\it{Planck}} $Y_{500}$ values calculated at the CARMA-8 cluster centroids using the CARMA-8-derived $\theta_{500}$ measurements. The ratio of $Y_{500}$ values from columns 2 and 4 is given in the fifth column, while the ratio of $\theta_{500}$ from the blind {\it{Planck}} (column 2) and the CARMA-8 analyses is provided in column 6 (where the CARMA-8 $\theta_{500}$ values have been taken from Table \ref{tab:paramsGNFW}. }
\label{tab:paramsY500}
\renewcommand{\arraystretch}{1.4}
\begin{tabular}{lcccccccccccc}
\hline \hline
Cluster ID & $Y_{500}$ blind & $\theta_{500}$ blind & $Y_{500}$ & $Y_{500, \rm{blind}}$/$Y_{500}$  & $\Theta_{500, \rm{blind}}$/$\theta_{500}$    \\
   & arcmin$^2$ & arcmin &  arcmin$^2$  & &  \\
   &  $\times 10^{-4}$ & & $\times 10^{-4}$ & \\  \hline
 P014 & 13.2 $\pm$ 6.4& 4.23 & $8.2_{-1.8}^{+1.9}$ & 1.61 & 0.98 \\
 P086 & 6.9 $\pm$ 2.5 & 4.23 & $5.8_{-1.7}^{+1.9}$ & 1.18 & 1.14  \\
 P097 &3.8 $\pm$ 0.8 & 0.92 & $3.1_{-0.5}^{+0.6}$ & 1.22 & 0.29 \\
 P109 & 5.4 $\pm$ 1.2 & 0.92  & $5.9_{-1.1}^{+1.5}$ & 0.91 & 0.30 \\
 P170 & 11.7 $\pm$ 4.1 & 4.75 & $7.1_{-1.2}^{+2.0}$ & 1.65  & 1.76 \\
 P187 & 11.5 $\pm$ 4.1 & 3.35 & $10.5_{-3.4}^{+3.7}$& 1.10 & 0.82 \\
 P190 & 7.9 $\pm$ 4.7 & 4.75 & $5.9_{-1.7}^{+1.9}$ & 1.33  & 1.36 \\
 P205 & 10.3 $\pm$ 4.9 & 3.35 & $10.7_{-3.8}^{+3.9}$ & 0.97 & 0.68 \\
  P351 & 14.2 $\pm$ 9.7 & 6.75 & $8.9_{-3.1}^{+2.9}$ & 1.60 & 1.27\\
\hline
\end{tabular}
\end{table*}

\begin{table*}
\caption{Estimated radio-source contamination in {\it{Planck}} for clusters in our sample. Column 2 provides the sum of all the 1.4-GHz `deconvolved' integrated flux-density measurements of NVSS sources detected within 5\arcmin\ of the {\it{Planck}} cluster centroid. Assuming a spectral index, $\alpha_{\rm{s}}$, of 0.72 (see Paper 1 for details), we extrapolated the NVSS flux densities to find the total flux density at 100 and 143\,GHz (columns 3 and 4), the most relevant {\it{Planck}} bands for SZ. The fifth column contains the (candidate) SZ decrement measured in the {\it{Planck}} 143-GHz band within its beam area in mJy. The last column presents the percentage radio-source contamination to the 143-GHz {\it{Planck}} candidate SZ decrement. This percentage is likely to be the amount by which the SZ flux is underestimated in the {\it{Planck}} data, since these are faint sources below the {\it{Planck}} point source detection limit.}
\label{tab:totflux}
  \tabcolsep=0.1cm
\begin{center}
\begin{tabular}{lrrrrr}
  \hline \hline
Cluster& $\Sigma$ Flux  & $\Sigma$ Flux & $\Sigma$ Flux &143-GHz {\it{Planck}}  & $\%$ Radio Source\\ 
       ID            &	densities			& densities			& densities			& SZ decrement & Contamination\\ 
        & at 1.4 GHz & at 100 GHz & at 143 GHz & inside Beam & to Planck SZ\\ \hline
   P014 & 131.20 & 6.10 & 4.70 & -77.3 & 6.1 \\ 
  P028 & 120.70 & 5.60 & 4.40 & -90.2 & 4.9 \\ 
  P031 & 14.10 & 0.60 & 0.50 & -57.6 & 0.9\\ 
  P049 & 62.60 & 2.90 & 2.30 & -55.7 & 4.1\\ 
  P052 & 293.00 & 13.60 & 10.50 & -67.6 &15.5\\ 
  P057 & 126.80 & 6.00 & 4.60 &-96.5 & 4.8\\ 
  P090 & 4.30 & 0.20 & 0.20 &-42.0 &0.5\\ 
  P097 & 5.90 & 0.20 & 0.20 & -38.4& 0.5\\ 
  P109 & 54.90 & 2.60 & 2.00 & -108.4& 1.8\\ 
  P121 & 31.30 & 1.50 & 1.10 & -79.9& 1.4\\ 
  P134 & 15.80 & 0.70 & 0.60 & -54.9& 1.1\\ 
  P138 & 37.00 & 1.70 & 1.30 & -66.8&1.9\\ 
  P170 & 18.40 & 0.90 & 0.70 & -141.1&0.5\\ 
  P187 & 71.10 & 3.40 & 2.60 & -70.4& 3.7\\ 
  P190 & 18.90 & 0.80 & 0.70 &-48.6 & 1.4\\ 
  P205 & 15.80 & 0.70 & 0.60 & -82.1 & 0.7\\ 
  P264 & 8.40 & 0.40 & 0.30 & -61.1&0.5\\ 
  P351 & 67.10 & 3.10 & 2.40 & -101.7&2.4\\ 
       \hline
\end{tabular}
\end{center}
\end{table*}

\subsection{Estimation of Radio-source Contamination in the {\it{Planck}} 143-GHz Data}
\label{sec:rsc}

In order to assess if there are any cluster-specific offsets in the {\it Planck} $Y_{500}$ values, we estimate the percentage of radio-source contamination to the {\it{Planck}} SZ decrement at 143\,GHz---an important {\it{Planck}} frequency band for cluster identification---from the 1.4-GHz NVSS catalog of radio sources.
Spectral indices between 1.4 and 31\,GHz were calculated in Table 3 in Paper 1 for sources detected in both our CARMA-8 LB data and in NVSS, giving a mean value of $\alpha_{\rm{s}}$ of 0.72. We use this value for $\alpha_{\rm{s}}$ to predict the source-flux densities at 100 and 143 GHz of all NVSS sources within 5\arcmin of the CARMA-8 pointing center,  following the same relation as we did earlier, $S_{\nu} \propto \nu^{-\alpha_{\rm{s}}}$.

The accuracy of the derived 100 and 143-GHz source fluxes is uncertain. Firstly, there is source variability due to the fact the NVSS and CARMA-8 data were not taken simultaneously, which could affect the 1.4-31\,GHz spectral index. Secondly, we assume the spectral index between 1.4 and 31\,GHz is the same as for 1.4 to 143\,GHz, which need not be true. Thirdly, we deduce $\alpha_{\rm{s}}$ from a small number of sources, all of which must be bright in the LB data and apply this $\alpha_{\rm{s}}$ to lower-flux sources found in the deeper NVSS data, for which $\alpha_{\rm{s}}$ might be different. However, previous work shows that this value for $\alpha_{\rm{s}}$ is not unreasonable. 
 Comparison of 31-GHz data with 1.4-GHz data on field sources has been previously done by \cite{muchovej2010} and \cite{mason2009}. For the former, the 1.4-to-31\,GHz spectral-index distribution peaked at 0.7 while, for the latter, it had a mean value of 0.7. The Muchovej et al. study also investigated the spectral index distribution between 5 and 31\,GHz and located its peak at $\approx 0.8$. Radio source properties in cluster fields have been characterized in e.g., \cite{coble2007} tend to have a steep spectrum. In particular, the 1.4-to-31\,GHz spectral index for the Coble et al. study had a mean value of 0.72. \cite{sayers2013} explored the 1.4-to-31\,GHz radio source spectral properties towards 45 massive cluster systems and obtained a median value for $\alpha_{\rm{s}}$ of 0.89, which they showed was consistent with the 30-to-140\,GHz spectral indices. The radio source population used to estimate the contamination to the Planck 143 GHz signal is likely to be a combination of field and cluster-bound radio sources due to the size of the Planck beam and the fact that some of the candidates might be spurious Planck detections. Overall, given the differences in the source selection and in frequency, and the agreement with other studies, our choice for of $\alpha_{\rm{s}} =0.72$ seems to be a reasonable one. 

In Table \ref{tab:totflux} we list the sum of all the predicted radio-source-flux densities at 100 and 143\,GHz of all the NVSS-detected sources within 5\arcmin\ of our pointing centre. This yields an approximate measure of the radio-source contamination in the {\it{Planck}} beam at these frequencies. The mean of the sum of all integrated source-flux densities at 1.4\,GHz is 61.0\,mJy (standard deviation, s.d. 71.6); at 100\,GHz it is 2.8\,mJy (s.d.$=3.3$) and at 143\,GHz it is 2.2\,mJy (s.d.=2.6). The SZ decrement towards each cluster candidate within the 143 GHz {\it{Planck}} beam is given in Table \ref{tab:totflux}, together with the (expected) percentage of radio-source contamination to the {\it{Planck}} cluster signal at this frequency, which on average amounts to $\approx 2.9\%$. The mean percentage contamination to the {\it{Planck}} SZ decrement would drop to $\approx 1.3\%$ if we used the \cite{sayers2013} $\alpha_{\rm{s}}=0.89$ and would increase to $\approx 5\%$ if we used a flatter $\alpha_{\rm{s}}$ of $0.6$. Thus, we expect the flux density from unresolved radio sources towards our cluster candidates to be an insignificant contribution to the {\it{Planck}} SZ flux although individual clusters may have radio source contamination at the $\sim5-15$\% level.

\begin{table*}
\centering
\small
\caption{Mean and 68\% confidence uncertainties for radio-source parameters for sources detected in the LB CARMA-8 data towards candidate {\it{Planck}} clusters with a CARMA-8 SZ detection. These parameters have been obtained from the joint cluster+sources fit in {\sc{McAdam}} to the SB data using cluster model I (Section \ref{sec:cm1}).}
\label{tab:sparams}
\begin{tabular}{lccccc}
\hline \hline 
   Source ID & Cluster ID        & ${\rm{RA}}_s$/deg     & ${\rm{Dec}}_s$/deg       & $S_{31}$/Jy       & $\alpha_{\rm{s}}$   \\ \hline
 1 & P014 & 240.831 $\pm$  0.001 &        3.282 $\pm$      0.001  & 0.0081 $\pm$  0.0012  & 0.5 $\pm$ 0.4 \\
 2 & P014 & 240.875 $\pm$ 0.001 & 344.155$\pm$  0.001 & 0.0089 $\pm$ 0.0022 & 0.6 $\pm$ 0.4 \\ 
 1 &P109	 & 275.718 $\pm$  0.002 &       78.384 $\pm$  0.002  & 0.0018 $\pm$  0.0004  & 0.6 $\pm$ 0.5\\
1 &P170		& 132.813 $\pm$  0.002 &       48.619 $\pm$  0.002  & 0.0046 $\pm$  0.0007  & 0.5 $\pm$ 0.5\\
1 &P187		& 113.084 $\pm$  0.001 &       31.688 $\pm$  0.001  & 0.0037 $\pm$  0.0004  & 0.5 $\pm$ 0.5\\
1 &P351 		& 226.077 $\pm$  0.002 &       -5.914 $\pm$  0.002  & 0.0028 $\pm$  0.0005  & 0.6 $\pm$ 0.5\\
 \hline
 \end{tabular}
\end{table*}

\subsection{Improved Constraints on $Y_{500}$ and $\theta_{500}$ from the Use of a {\it{Planck}} Prior on $Y_{500}$ in the Analysis of CARMA-8 Data}
\label{sec:joint}

Due to their higher resolution (a factor of $\gtrsim 5$), the CARMA-8 data are better suited than the {\it{Planck}} data to constrain $\theta_{500}$. On the other hand, the large {\it{Planck}} beam (FWHM$\approx 10\arcmin$ at 100\,GHz) allows the sampling parameter $Y_{\rm{T}}$ for our clusters (all of which have $\theta_{500} \lesssim 5\arcmin$) to be measured directly, which is not the case for the CARMA-8 data due to its finite sampling of the {\it{uv}} plane and the missing zero-spacing information (a feature of all interferometers). We have exploited this complementarity of the {\it{Planck}} and CARMA-8 data to reduce uncertainties in $Y_{500}$ and $\theta_{500}$. In order to do this, we filtered out the parameter chains (henceforth chains) for the analysis of the CARMA-8 data (model I) that had values of $Y_{500}$ outside the range allowed by the {\it{Planck}} $Y_{500}$ results (Table \ref{tab:paramsY500}). We refer to the results from the remaining set of chains as the {\it{joint}} results (Table \ref{tab:paramsY500joint}). 
In Figure \ref{fig:combi} we plot the 2D marginalized distributions for $Y_{500}$ and $\theta_{500}$ for the CARMA-8 data alone (black contours) and for the joint results (magenta contours). Similar approaches comparing {\it{Planck}} data with higher resolution SZ data have been undertaken by \cite{PlanckAMI} (with AMI), \cite{muchovej2012} (with CARMA), \cite{sayers2013} (with BOLOCAM) and \cite{perrott2014} (with AMI). 
Clearly, the introduction of cluster size constraints from high resolution interferometry data provides a powerful way to shrink the uncertainties in $Y_{500}-\theta_{500}$ phase space.

\begin{table}
\centering
\small
\caption{Bayes Factor, $K$, for SZ signals detected in the CARMA-8 data by McAdam. Since the prior ratio is set to unity, the Bayes Factor provides a measure of the quality of the model fit to the data.
For two clusters, P014 and P134, two potential SZ signals were detected in the Field of View (FoV) of each observation; the second cluster-like signal is labelled with a `b'. We adopt Jeffreys (1961) interpretation of $K$, though with fewer categories, to be even more conservative. We consider $K\leq 0.1$ to be strong evidence against a cluster signal (NC); $0.1 < K \leq10 $ means our data cannot be used on their own to distinguish robustly between a model with or without a cluster signal (ND) and $K > 10$ indicates there is strong evidence for the presence of a cluster signal in the data (D). For reference, the signal-to-noise ratio (SNR) for detected clusters (those highlighted in bold font) in CARMA-8 and {\it{Planck}} data are given in Table 2 of Paper 1.}
\label{tab:evidences}
\begin{tabular}{lccc}
\hline \hline
 Cluster & Bayes Factor & Degree of detection\\ 
  \hline
 {\bf{P014}} & 3.4e+01 & D \\ 
   P014b & 4.2e-01& ND\\
   P028 & 4.6e-01 & ND \\ 
  P031 & 5.0e-02 & NC \\ 
 P049 & 2.2e-01 & ND \\ 
 P052 & 9.5e-01 & ND \\ 
 P057 & 6.1e-01 & ND \\ 
 {\bf{P086}} & 3.5e+03 & D \\ 
 P090 & 1.2e+00 & ND \\ 
  {\bf{P097}} & 7.9e+02 & D \\ 
 {\bf{P109}} & 8.4e+02 & D \\ 
 P121 & 4.9e-01 & ND \\ 
 P134 & 3.0e-02 & NC \\ 
  P134b & 3.0e-04 & NC\\
  {\bf{P170}} & 7.2e+08 & D \\ 
  {\bf{P187}} & 8.0e+08 & D \\ 
   {\bf{P190}} & 5.6e+18 & D \\ 
   {\bf{P205}} & 1.3e+09 & D \\ 
  P264 & 2.3e+00 & ND \\ 
  {\bf{P351}} & 2.7e+01 & D \\ 
   \hline
\end{tabular}
\end{table}

\section{Discussion}
\label{sec:disc}

\subsection{Use of Priors}

When undertaking a Bayesian analysis, it is important not only to check that the priors on individual parameters are sufficiently wide, such that the distributions are not being truncated, but also that the {\it{effective prior}} is not biasing the cluster parameter results. Here the term {\it{effective prior}} refers to the prior that is being placed on a model parameter while taking into account the combined effect from all the priors given to the set of sampling parameters. What may seem to be  inconspicuous priors on individual parameters can occasionally jointly re-shape the high dimensional parameter space in unphysical ways; this was noticed in e.g., \cite{zwart2011}. Biases from effective priors should be investigated by undertaking the analysis without data i.e., by setting the likelihood function to a constant value. Such studies for the models used in this work have been  presented in \cite{carmen2012},  \cite{olamaie2012} and \cite{olamaie2013} and have determined that the combination of all the model priors does not bias the results.

\begin{table}
\centering
\small
\caption{Columns 2 and 3 contains the joint $\theta_{500}$ and $Y_{500}$ values, which were computed by truncating the output of the CARMA-8 chains to have $Y_{500}$ values in the range allowed by the {\it{Planck}} results (column 4 in Table \ref{tab:paramsY500}).}
\label{tab:paramsY500joint}
\renewcommand{\arraystretch}{1.4}
\begin{tabular}{lcccc}
\hline \hline
Cluster ID & $\Theta_{500, \rm{Joint}}$  & $Y_{500, \rm{Joint}}$  \\
   & arcmin &  arcmin$^2$ \\
   &    &  $\times 10^{-4}$ \\  \hline
 P014 &$3.8^{+ 0.1}_{-0.3 }$  &$ 8.2^{+ 0.6}_{- 0.7}$\\
 P086  &$2.7^{+ 0.3}_{-0.4 }$  &$ 5.5^{+ 0.5 }_{-0.7}$\\
 P097 &$2.1^{+ 0.1}_{-0.3 }$  &$ 3.3^{+ 0.2 }_{-0.2}$\\
 P109 &$3.2^{+ 0.2}_{-0.4 }$  &$ 5.7^{+ 0.3 }_{-0.5}$\\
 P170 & $3.3^{+ 0.2}_{-0.2}$  &$ 7.1^{+ 0.4}_{-0.6}$\\
 P187 &$4.2^{+ 0.3}_{-0.4 }$  &$  10^{+ 1.0}_{-1.4}$\\
 P190 &$2.8^{+ 0.2}_{-0.2 }$  &$ 5.8^{+ 0.5}_{-0.7}$\\
 P205 &$4.3^{+ 0.3}_{-0.5 }$  &$  10^{+1.1}_{-1.5}$\\
  P351& $5.5^{+0.2}_{-0.8 }$  &$  8^{+ 0.7}_{-1.2}$\\
\hline
\end{tabular}
\end{table}
 
\subsection{Characterization of the Cluster Candidates}
 
\subsubsection{Cluster position and Morphology}

The mean separation (and standard deviation, s.d.) of the CARMA-8 centroids from Model I and the {\it{Planck}} position is $\approx 1.5\arcmin$ (0.5); see Tables \ref{tab:paramsGNFW} and \ref{tab:paramsBLOB} for offsets from the CARMA-8 SZ decrement to the {\it{Planck}} position. This offset is  comparable to the offsets between {\sc{Planck}} and X-ray cluster centroids found for the ESZ (\citealt{ESZ}) and the PSZ (\citealt{Planck2013}), which were typically $\approx 2\arcmin$ and $70\arcsec$, respectively. The cluster candidate with the largest separation, $\approx 2.5\arcmin$, is P014. The high-resolution CARMA-8 data allow for the reduction of positional uncertainties in the {\it{Planck}} catalog for candidate clusters from a few arcminutes to within $\lesssim 30''$. This is crucial, amongst other things, for the efficient follow-up of these candidate systems at other wavelengths. P351 has the largest positional uncertainties for both parameterizations ($\approx 40\arcsec$), indicative of a poorer fit of the models to the data, since the noise in the CARMA-8 data is one of the smallest of the sample. Interestingly, this cluster stands out in the $\beta$ parameterization for having the shallowest profile (Figure  \ref{fig:tprofile}), and in the gNFW parameterization for having the largest $\theta_{500}$. 
Overall, the positional uncertainties from the shape-fitting model I, tends to be larger than that from the radial profile based model II,
typically by a factor of 1.2 and reaching a factor of 2.8. The different parameter degeneracies resulting from each analysis is likely to be the dominant cause for this. As shown in Figure 1 of \cite{carmen2012}, in the $Y_{500}-\theta_{500}$ plane, the 2D marginalized distribution for the cluster size is significantly narrower for the $\beta$ parameterization (model II)  than for the gNFW parameterization (model I).

\begin{figure}
\begin{center}
\centerline{\includegraphics[width=8cm,height=8cm,clip=,angle=0.]{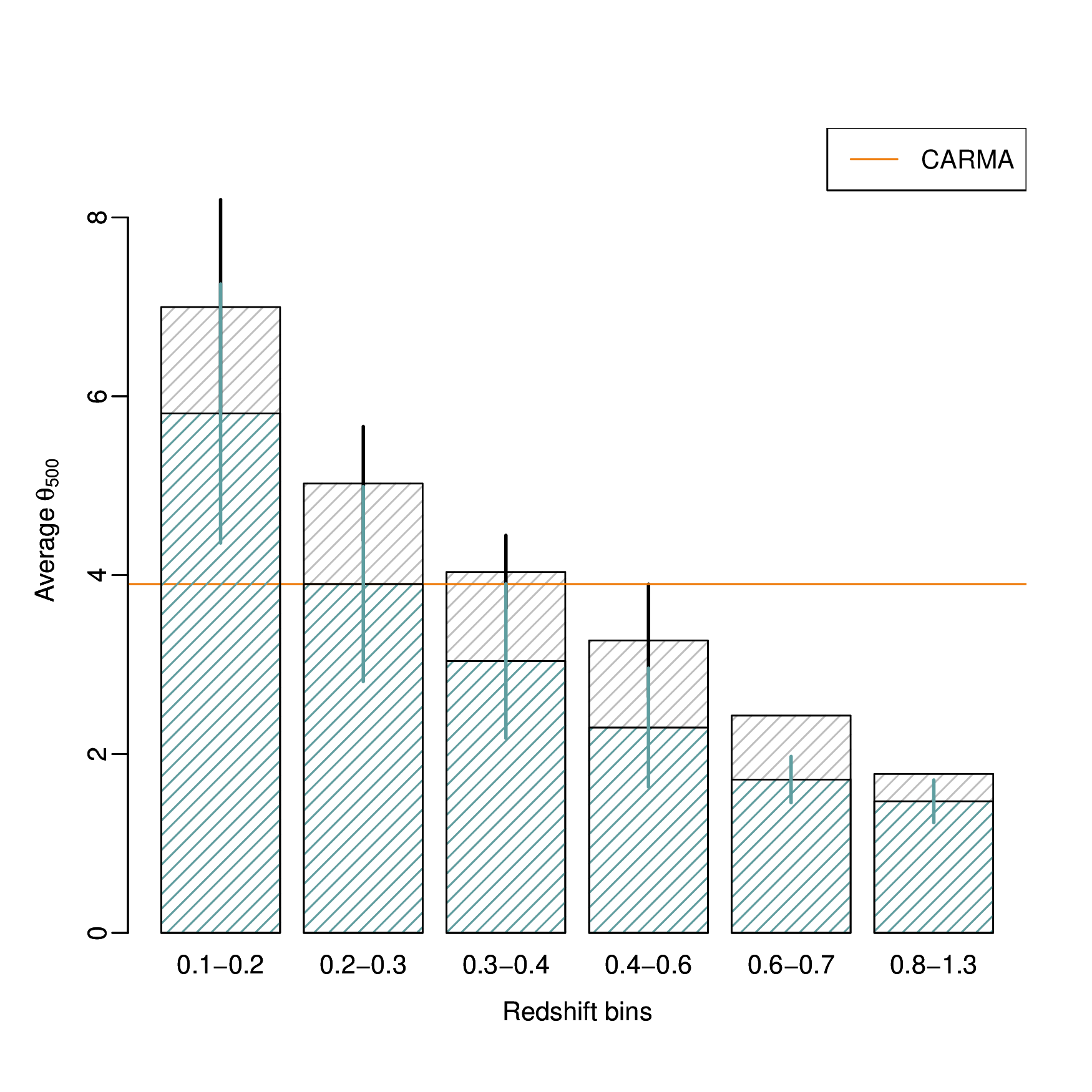}}
\caption{Average $\theta_{500}$,  $\left<\theta_{500}\right>$,  values calculated from the MCXC catalog \protect\citep{pifa2010} within a a series of redshift bins, as indicated in the x-axis e.g., the first bin represents  $\left<\theta_{500}\right>$  for MCXC clusters with $0.1 \leq z < 0.2$. 
The barplots filled with grey diagonal lines use all the relevant MCXC cluster entries to get $\left<\theta_{500}\right>$, while those filled with blue diagonal lines, only include the more massive clusters, $M_{500}>4\times 10^{14}M_{\odot}$, which we expect to be more representative of our target sample as {\it{Planck}} detects the most massive clusters. Standard deviations are displayed as vertical lines. The average CARMA-8-derived $\theta_{500}$ estimate for our clusters is depicted with an orange horizontal line. This plot suggests the typical $\theta_{500}$ values for our clusters are most comparable with the $\theta_{500}$ values for MCXC clusters at $0.2 \lesssim z \lesssim 0.4$. Furthermore, when combined with our photometric redshift
measurements in Paper 1 which place our clusters at $z\sim0.5$, the data suggest that we are finding the largest, most-massive, clusters at intermediate redshifts.
}
\label{fig:thetaMCXC}
\end{center}
\end{figure}

\begin{figure}
\begin{center}
\centerline{\includegraphics[width=8cm,height=8cm,clip=,angle=0.]{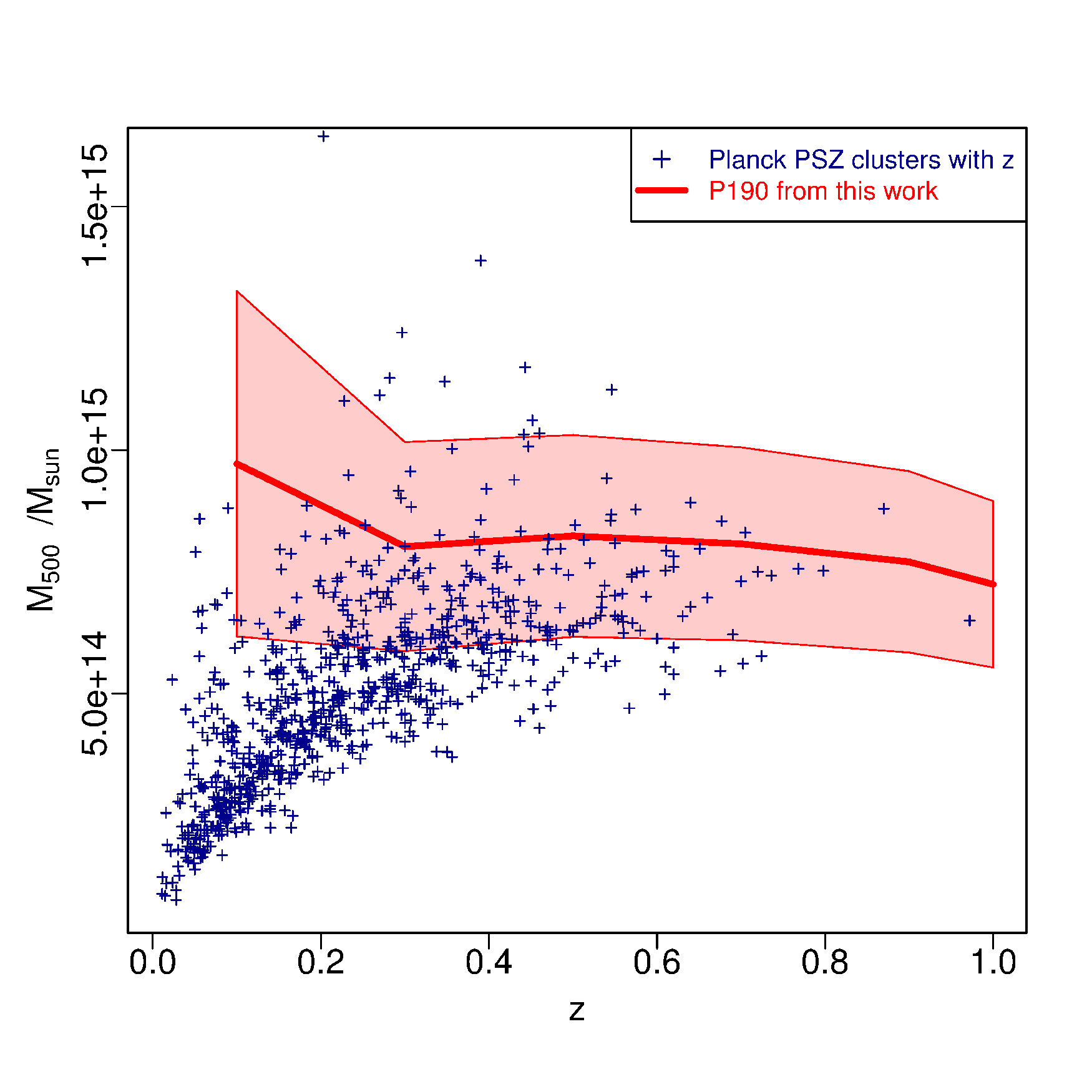}}
\caption{$M_{500}$ estimates for candidate cluster P190 as a function of redshift. To obtain the $M_{500}$ measurements a third parameterization, model III (Section \ref{sec:mass}; \citealt{olamaie2013}), which samples directly from the cluster mass, was implemented in the McAdam analysis. Since this model requires redshift information as an input, but spectroscopic-redshift information is not available for most of our candidate clusters, we ran this model run six times, from $z=0.1$ to $z=1.0$ in steps of 0.2. The range of $M_{500}$ values encompassing 68$\%$ of the probability distribution are shown in the red shaded area. We overplot the $M_{500}$ estimates from the {\it{Planck}} PSZ cluster catalog (\citealt{Planck2013}) for entries with an associated redshift in blue + signs. The PSZ catalog of clusters contains the majority of the most massive X-ray-detected clusters in the MCXC catalog. Hence, the clusters in our sample, which have SZ signatures similar to that of P190, are amongst the most massive known systems at intermediate redshifts.}
\label{fig:mass}
\end{center}
\end{figure}

On average, cluster candidates with CARMA-8 detections have $\theta_{s}=3.4 \pm 0.2'$ with an s.d. of $0.9$ and $\theta_{500}= 3.9 \pm 1.6'$, with s.d. of 0.9 (see Table \ref{tab:paramsGNFW}). The largest cluster has $\theta_{s}=5'$, $\theta_{500}=5.3'$  (P351) and the smallest $\theta_s= 2' $, $\theta_{500}=2.7'$ (P170). 
In Paper 1 we estimated the photometric redshifts for our cluster candidates with a CARMA-8 SZ detection and found that, on average, they appeared to be at $z\approx0.5$; in Section \ref{sec:specz} we report on the spectroscopic confirmation of P097 at $z=0.565$. The relatively small values for $\theta_{500}$ would support the notion that our systems are at intermediate redshifts ($z \gtrsim 0.5$). In comparison, for the MCXC catalog of X-ray-identified clusters (\citealt{pifa2010}), whose mean redshift is 0.18, the mean X-ray-derived $\theta_{500}$ is a factor of 2 larger. In Figure \ref{fig:thetaMCXC} we plot the average $\theta_{500}$ within a series of redshift ranges starting from $z=0.1$ for all MCXC clusters (in grey) and for only the more massive, $M_{500}>4\times 10^{14}M_{\odot}$, clusters (in blue), which should be more representative of the cluster candidates analyzed here (see Figure \ref{fig:mass}) and mark the average CARMA-8-derived $\theta_{500}$ for our clusters with an orange line. This plot suggests the $\theta_{500}$ values for our clusters are most comparable with the $\theta_{500}$ values for MCXC clusters at $0.2 \lesssim z \lesssim 0.4$.

The resolution of the CARMA-8 data, together with the often poor signal-to-noise ratios and complications in the analysis, e.g., regarding the presence radio sources towards some systems, makes getting accurate measurements of the ellipticity $\eta$ challenging, with typical uncertainties in $\eta$ of 0.2 (Table \ref{tab:paramsGNFW}). The mean and standard deviation of $\eta$ for our sample is $0.7 \pm 0.07$. The values are therefore consistent with unity, 
to within the uncertainties. Nevertheless, the use of a spherical model is physically motivated and allows the propagation of realistic sources of uncertainty. Moreover, comparison of models with spherical and elliptical geometries for similar data from AMI is presented in \cite{tash2012} which show the Bayesian evidences are too alike for model comparison, indicating that the addition of complexity to the model by introducing an ellipticity parameter is not significantly penalised. In \cite{perrott2014} modelling of AMI cluster data with and ellipsoidal GNFW profile instead of a spherical profile had a negligible effect on the constraints in $Y$. Our CARMA data with higher noise levels and generally more benign source environments should show even smaller effects.

$\eta$ values close to 1 would be expected for relaxed systems, whose projected signal is close to spheroidal, unless the main merger axis is along the line of sight. On the other hand, disturbed clusters should have $\eta \rightarrow 0.5$. Some evidence for a correlation of cluster ellipticity and dynamical state has been found in simulations, e.g, \cite{krause2012} and data, e.g., \cite{kolo2001} (X-ray), \cite{plionis2002} (X-ray and optical) and  \cite{carmen2012} (SZ), although this correlation has a large scatter.
Hence, from the derived fits to the data, we conclude that our sample is likely to be mostly comprised by large, dynamically active systems, unlikely to have fully virialized which is not surprising given the intermediate
redshifts of the sample. 

\begin{figure}
\begin{center}
\centerline{\includegraphics[width=8.0cm,clip=,angle=0.]{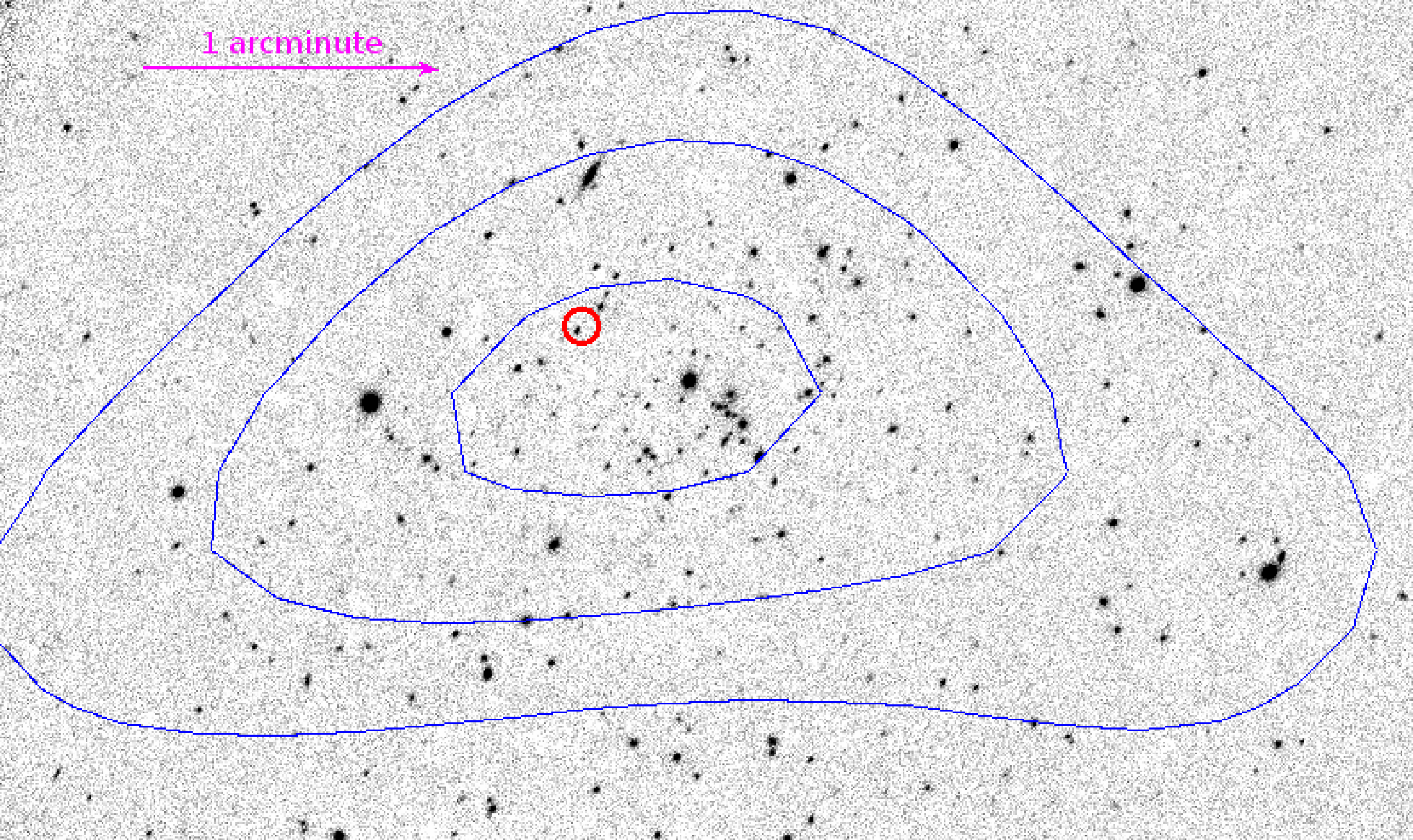}}
\caption{Keck/MOSFIRE Y-band image for P097 in grey background, where high intensity regions appear darker. Overlaid are the CARMA SZ contours, as shown in Paper I, in blue. The contours are  2, 3 and 4$\sigma$, where $\sigma$, the RMS of the short baseline CARMA data towards P097, is $0.653$\,mJy/beam. The red circle shows the galaxy the spectroscopic target whose redshift was determined to be 0.565}
\label{fig:keck}
\end{center}
\end{figure}

\subsection{Cluster-Mass Estimate}
\label{sec:mass}

To estimate the total cluster mass $M_{500}$ within $r_{500}$, we use the \cite{olamaie2013} cluster parameterization, which samples directly from $M_{200}$\footnote{$M_{500}$ is determined by calculating $r_{500}$, which in turn is computed by equating the expression for mass from the NFW density profile within $r_{500}$ and the mass within an spherical volume of radius $r_{500}$ under the assumption of spherical geometry.
Under this NFW-gNFW-based cluster parameterization the relation between $M_{200}$ and $M_{500}$ is $M_{200}=1.35 \times M_{500}$. For further information on this cluster model see \protect\cite{olamaie2013}.}. This model describes the cluster dark matter halo with an NFW profile (\citealt{navarro1996}) and the pressure profile with a gNFW profile (\citealt{nagai2007}), using the set of gNFW parameters derived by \cite{arnaud2009}. There are two other additional assumptions of this model: (1) the gas is in hydrostatic equilibrium and (2) the gas mass fraction is small compared to unity ($\lesssim 0.15$). \footnote{{Extensive work has been done to characterize the hydrostatic mass bias, which has been shown to range between $\approx 5$ and $45\%$ depending on dynamical state in numerical simulations (e.g., \citealt{rasia2006}, \citealt{nagai2007b} and \citealt{molnar2010}) and studies comparing SZ and weak lensing mass estimates (e.g., \citealt{zhang2010} and \citealt{mahdavi2013}).  }} The cluster redshift is a necessary input to this parameterization. In the absence of spectroscopic data towards our cluster candidates to get an accurate redshift estimate\footnote{P187 might an exception as the observational evidence shows that it is likely to be associated with a well-known Abell cluster, Abell\,586, at $z=0.171$. We obtained spectroscopic confirmation for P097, see Section \ref{sec:specz}}, coarse photometric redshifts based on SDSS and WISE colors were calculated in Paper 1. In Figure \ref{fig:mass} we plot the $M_{500}$ estimate (mean values are depicted by the thick line and the area covering the 68$\%$ of the probability distribution has been shaded in red) as a function of $z$ for one of our cluster candidates, P190. To produce this plot we ran the \cite{olamaie2013} cluster parameterization  six times  using a Delta-prior on redshift,  which we set to values from $0.1$ to $1.0$ in steps of 0.2. We chose P190 since it is quite representative of our cluster candidates and, at $8\sigma$, where $\sigma$ is the short baseline rms, it has the best SNR of the sample (see Table 2 in Paper 1). The photometric-redshift estimate for P190 from Paper 1 was $0.5$, which is also the average expected photometric redshift for the sample of CARMA-8 SZ detections. At this redshift, $M_{500}=0.8\pm0.2 \times10^{15}M_{\odot}$. As seen in Figure \ref{fig:mass}, after $z\approx0.3$ $M_{500}$ is a fairly flat function of $z$ (since the $z$ dependence in the model is carried by the angular diameter distance), such that, to one significant figure, our mean value is identical.

\subsubsection{Spectroscopic Redshift Determination for P097}
\label{sec:specz}

We have measured the spectroscopic redshift of a likely galaxy member
of P097 through Keck/MOSFIRE Y-band spectroscopy. We deem it a likely member, given that it is situated close to the peak of the CARMA SZ
decrement (within the $4\sigma$ contour) and close to a group of tightly clustered galaxies, as shown in Figure \ref{fig:keck}, of similar colours.
We detect clear evidence for H$\alpha$, [NII], [SII] at redshifts of 0.565, with 
the strongest lines being present in a Sloan-detected galaxy located at 
RA,DEC of (14:55:25.3,+58:52:33.86; J2000). We also find evidence
of velocity structure in this galaxy with the  H$\alpha$ being double
peaked with the line components separated by $\sim$150 km/s. These
data will be published in a separate paper.
We calculated $M_{500}$ for this cluster, as we did for P190 in the previous section, setting the redshift prior to a delta function at $z=0.565$ and
obtained $M_{500}=0.7\pm 0.2 \times 10^{15}M_{\odot}$, supporting the notion that our sample of clusters are some of the most massive
 clusters at $z\gtrsim 0.5$. Further follow-up of this sample in the X-rays and through weak lensing measurements with {\it Euclid} will help constrain
the mass of these clusters more strongly.

 \begin{table}
\caption{This table shows the average improvement in the SNR of the $Y_{500}$ measurements taken from the independent analysis of (blind) {\it{Planck}} or CARMA-8  data versus the joint analysis of {\it{Planck}} and CARMA-8 data. For clarity in the table we use a shorthand notation. We refer to ${\rm{SNR}}_{Y_{500, {\rm{Joint}}}}$ as the ratio of $Y_{500, {\rm{Joint}}}$ and the 68$\%$ error in $Y_{500, {\rm{Joint}}}$, that is, $Y_{500, {\rm{Joint}}}/\sigma Y_{500, {\rm{Joint}}}$, where, as denoted by the suffix label, the results are obtained from the joint analysis of the CARMA-8 and {\it{Planck}} data. Similarly, ${\rm{SNR}}_{Y_{500, {\rm{blind}}}}$ and ${\rm{SNR}}_{Y_{500, {\rm{CARMA-8}}}}$ are the ratios of  $Y_{500}$  and the 68$\%$ error in $Y_{500}$ for the independent analysis of {\it{Planck}} (blind) data and CARMA-8 data, respectively. Since we have asymmetric errors for $Y_{500}$, in the first column we provide the SNR ratios calculated using the lower-bound errors and in the second column those calculated using the upper-bound errors. In an entirely analogous manner, we calculate the improvements in the SNR of $\theta_{500}$ as we shift from independent analyses of the {\it{Planck}} or CARMA-8 data to a joint analysis. In the lower sub-table, we focus on the improvements to the 68$\%$-confidence uncertainties in $Y_{500}$ and $\theta_{500}$ from undertaking joint {\it{Planck}} and CARMA-8 analyses rather than using each data set on its own.
It is clear that a joint analysis of the two datasets provides much tighter constraints in the $Y_{500}-\theta_{500}$ parameter space. This is mostly due to the challenge in constraining cluster sizes in the {\it Planck} data and a lack of sensitivity to the signal from the outskirts of the cluster in the CARMA-8 data. }
\label{tab:snrimp}
{\renewcommand{\arraystretch}{0.5}} 
\begin{center}
\begin{tabular}{lrrrrr}
  \hline \hline
$\left< \frac{{\rm{SNR}}_{Y_{500, {\rm{Joint}}}}}{{\rm{SNR}}_{Y_{500, {\rm{CARMA-8}}}}}\right >$ & 4.9 & 6.5 \\
  $\left <\frac{{\rm{SNR}}_{\Theta_{500, {\rm{Joint}}}}}{{\rm{SNR}}_{\Theta_{500, {\rm{CARMA-8}}}}}\right >$  & 6.6  & 4.0 \\
  $\left <\frac{{\rm{SNR}}_{Y_{500, {\rm{Joint}}}}}{{\rm{SNR}}_{Y_{500, {\rm{blind}}}}}\right >$ & 5.2  & 3.8 \\
  $\left <\frac{{\rm{SNR}}_{\Theta_{500, {\rm{blind}}}}}{{\rm{SNR}}_{\Theta_{500, {\rm{Joint}}}}}\right >$ &  75.9  & 36.5 \\  \hline  
  & & & \\
    $\left< \frac{\sigma Y_{500, {\rm{CARMA-8}}}}{ \sigma Y_{500, {\rm{Joint}}}} \right>$ & 6.9 & 9.4\\
  $\left< \frac{\sigma \theta_{500, {\rm{CARMA-8}}}}{ \sigma \theta_{500, {\rm{Joint}}}} \right>$ & 7.3 & 4.4 \\
    $\left< \frac{\sigma Y_{500, {\rm{blind}}}}{ \sigma Y_{500, {\rm{Joint}}}} \right>$ & 7.3 & 5.2 \\
  $\left< \frac{ \sigma \theta_{500, {\rm{Joint}}}} {\sigma \theta_{500, {\rm{blind}}}}\right>$ & 52.2 & 27.7 \\ \hline
\end{tabular}
\end{center}
\end{table}

\addtolength{\subfigcapskip}{0pt}
\addtolength{\subfigtopskip}{-35pt}

\setcounter{subfigure}{0}
\begin{figure*}
\begin{center}
\setlength{\tabcolsep}{1mm}
\subfigure[Left to right: P014, P086, P097]{
\centering
\vspace{-2\baselineskip}
\includegraphics[width=6.5cm, height=6.5cm,trim=45 48 60 48,clip= ,angle=0.]{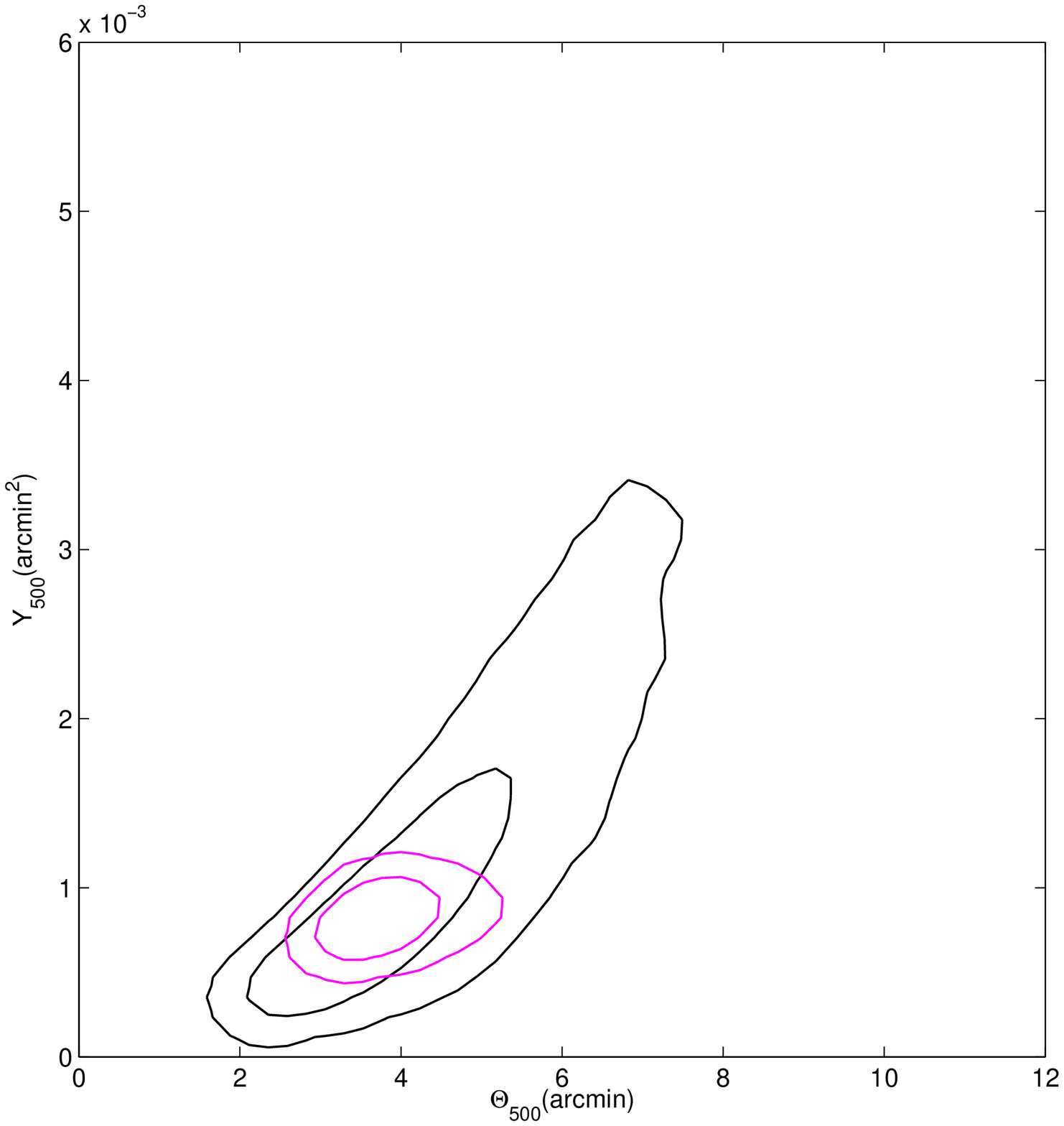}\hspace*{0.00001em}\includegraphics[width=6.5cm, height=6.5cm,trim=60 48 60 48,clip=,angle=0.]{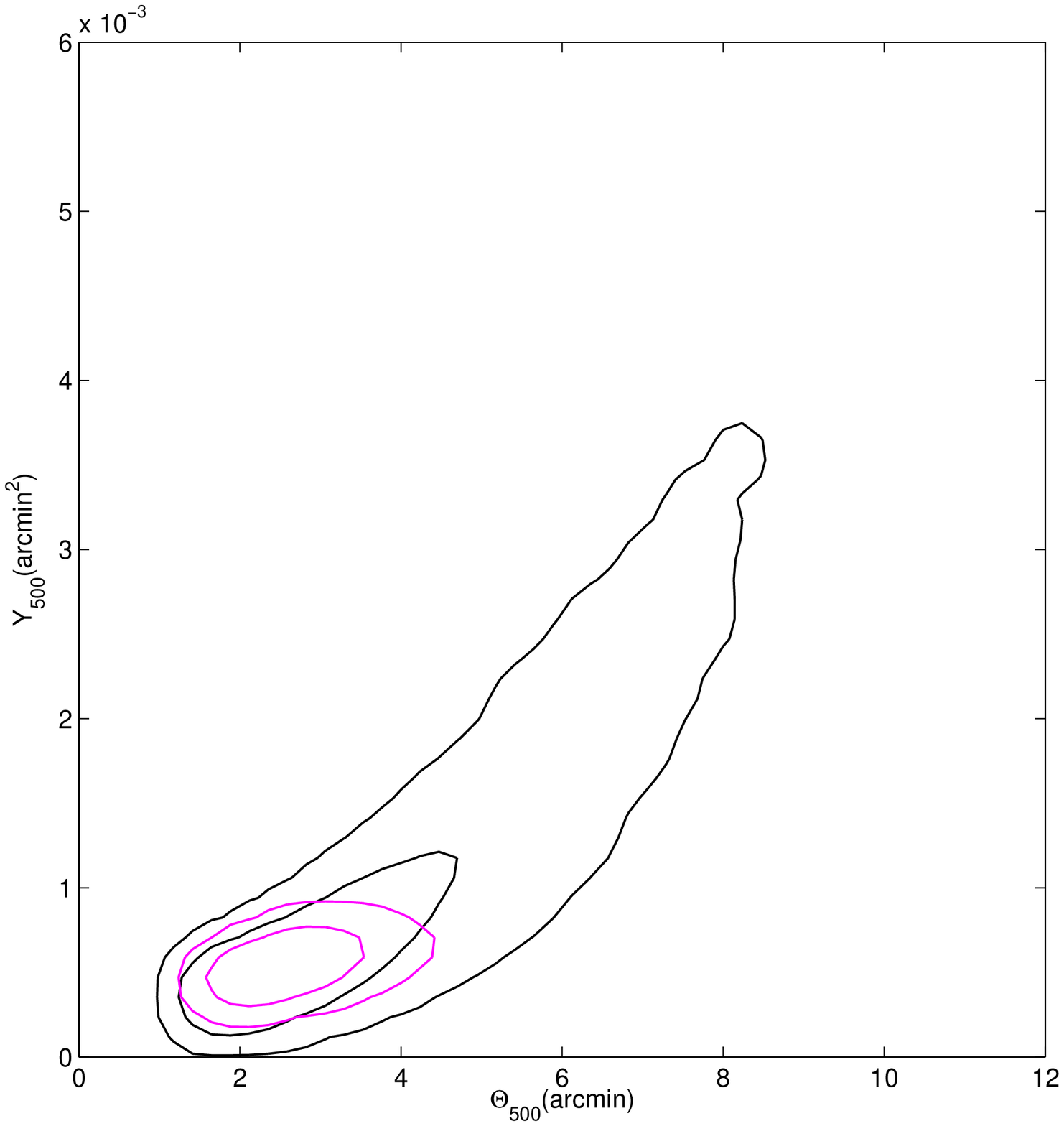}\hspace{0.05em}\includegraphics[width=6.5cm, height=6.5cm,trim=60 48 60 48,clip=,angle=0.]{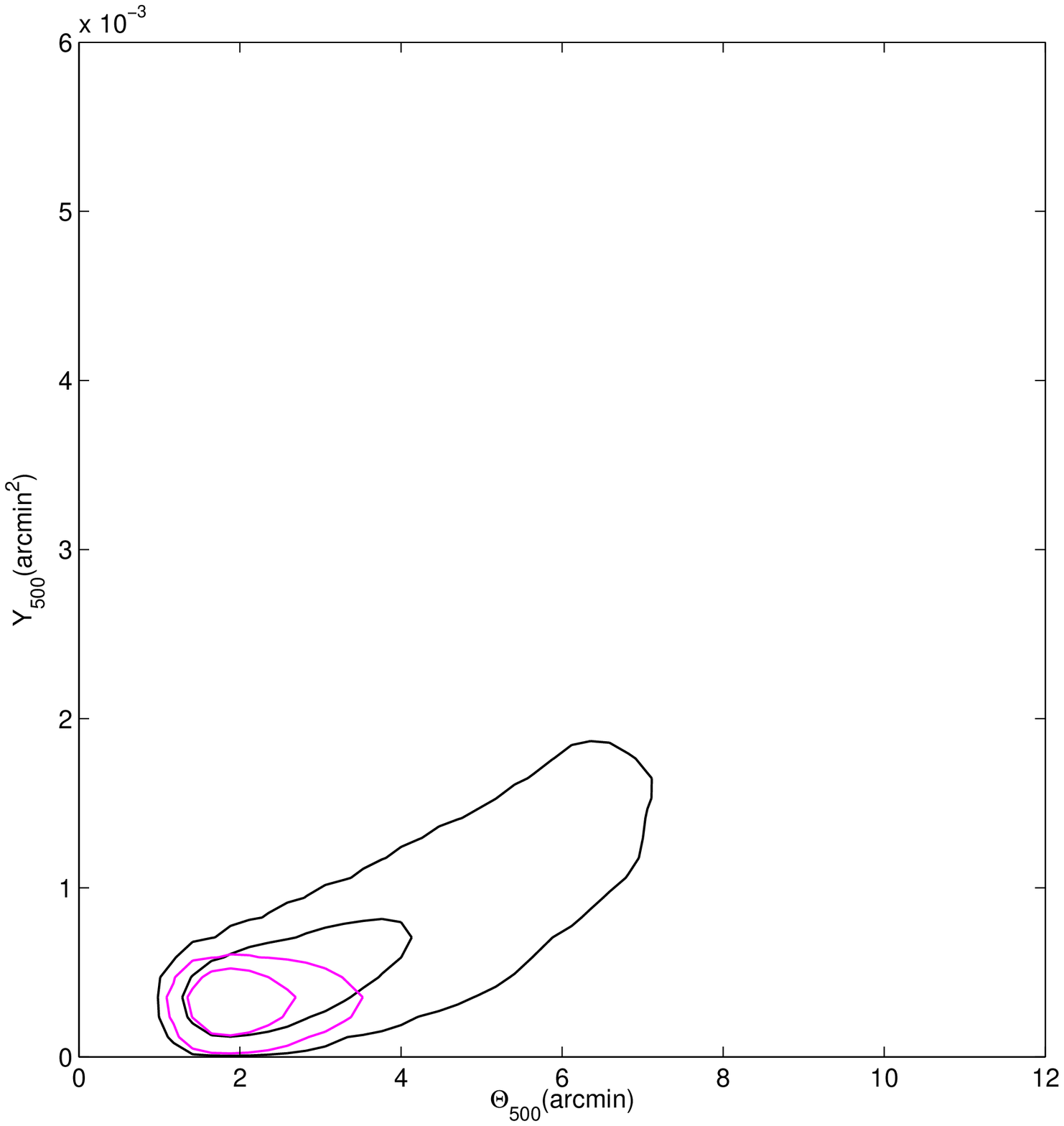}}\\
\subfigure[Left to right:   P109, P170, P187]{
\centering
\includegraphics[width=6.5cm, height=6.5cm,trim=60 48 60 48,clip= ,angle=0.]{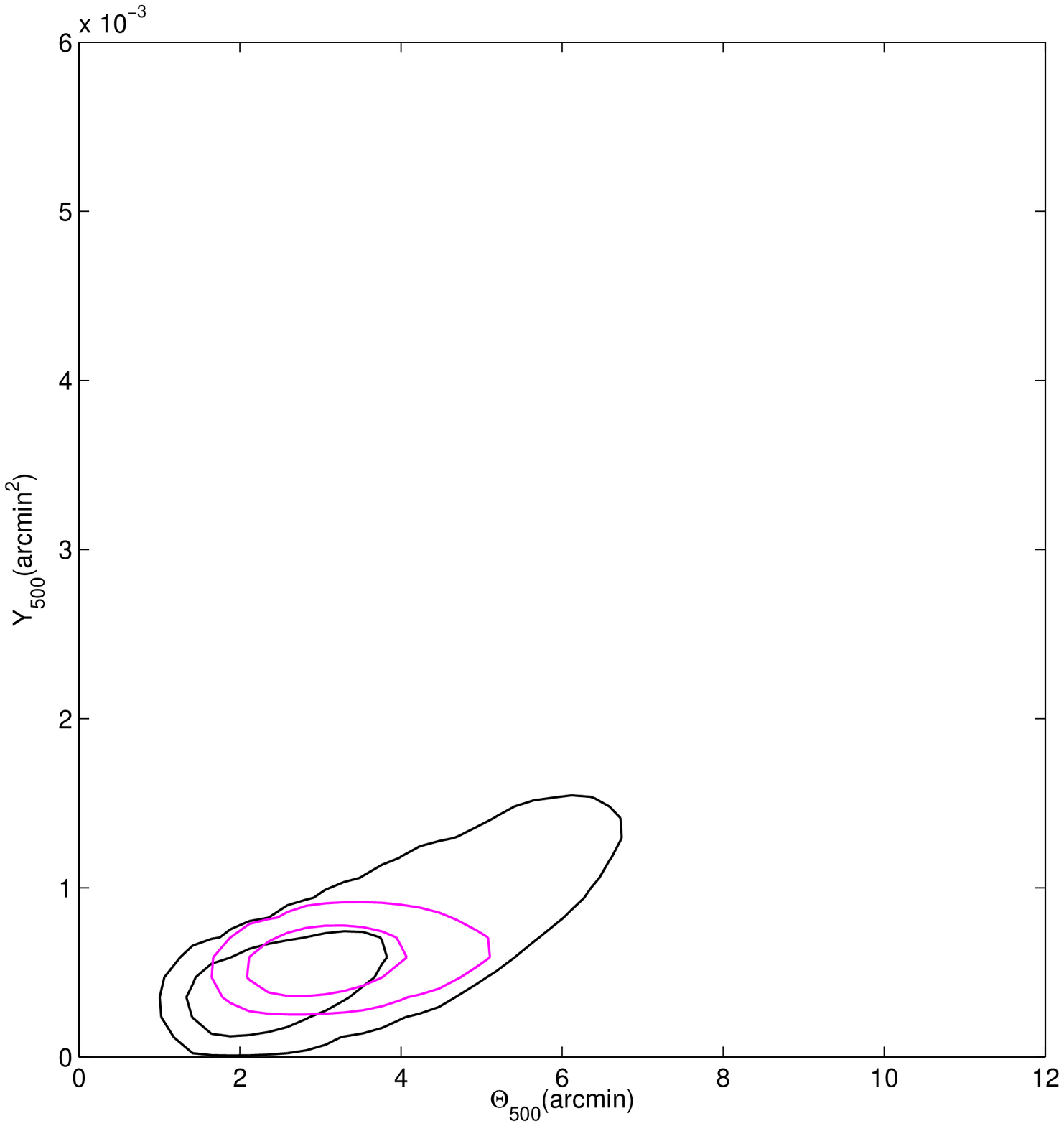}\hspace{0.05em}\includegraphics[width=6.5cm, height=6.5cm,trim=60 48 60 48,clip=,angle=0.]{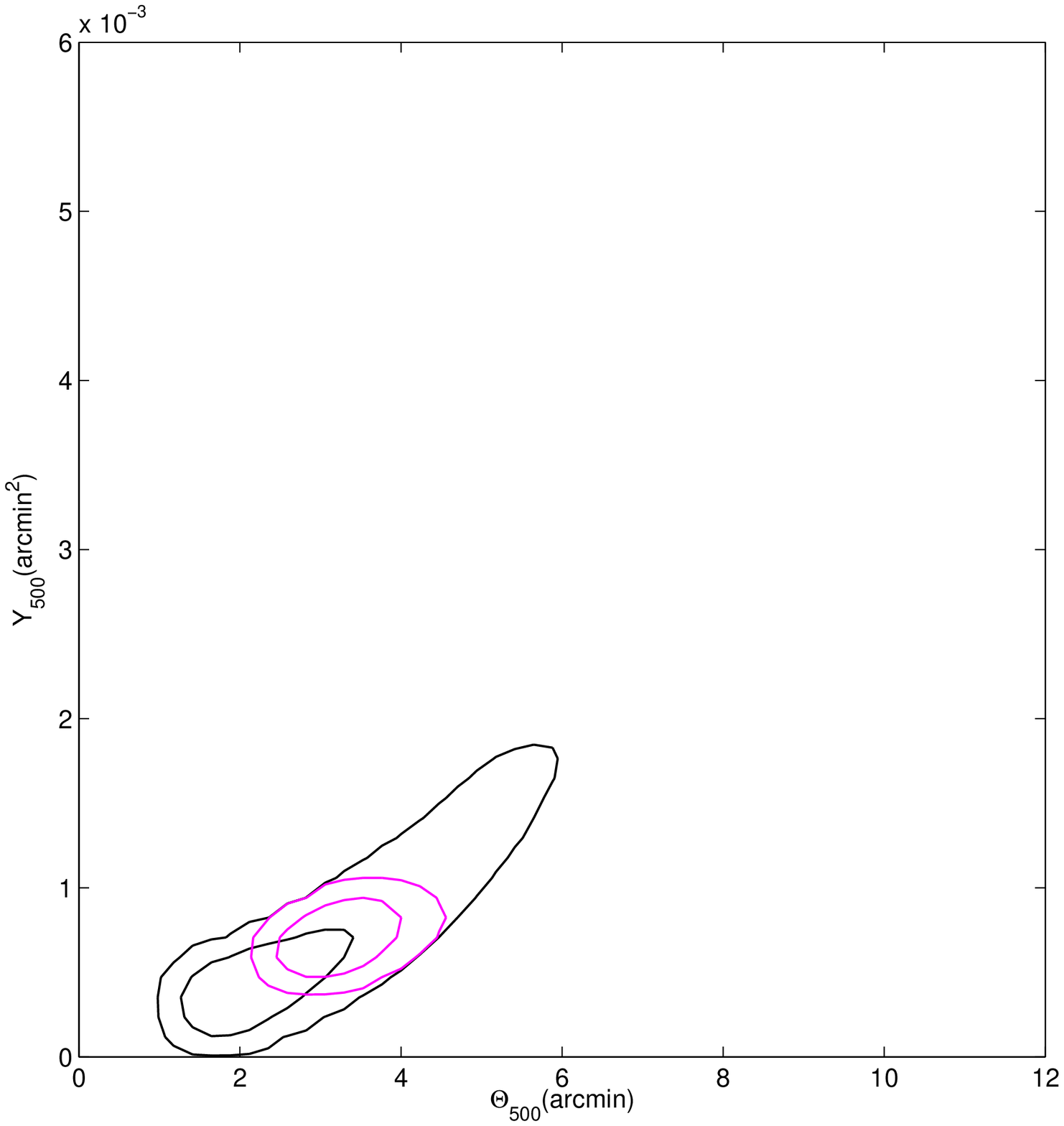}\hspace{0.05em}\includegraphics[width=6.5cm, height=6.5cm,trim=60 48 60 48,clip=,angle=0.]{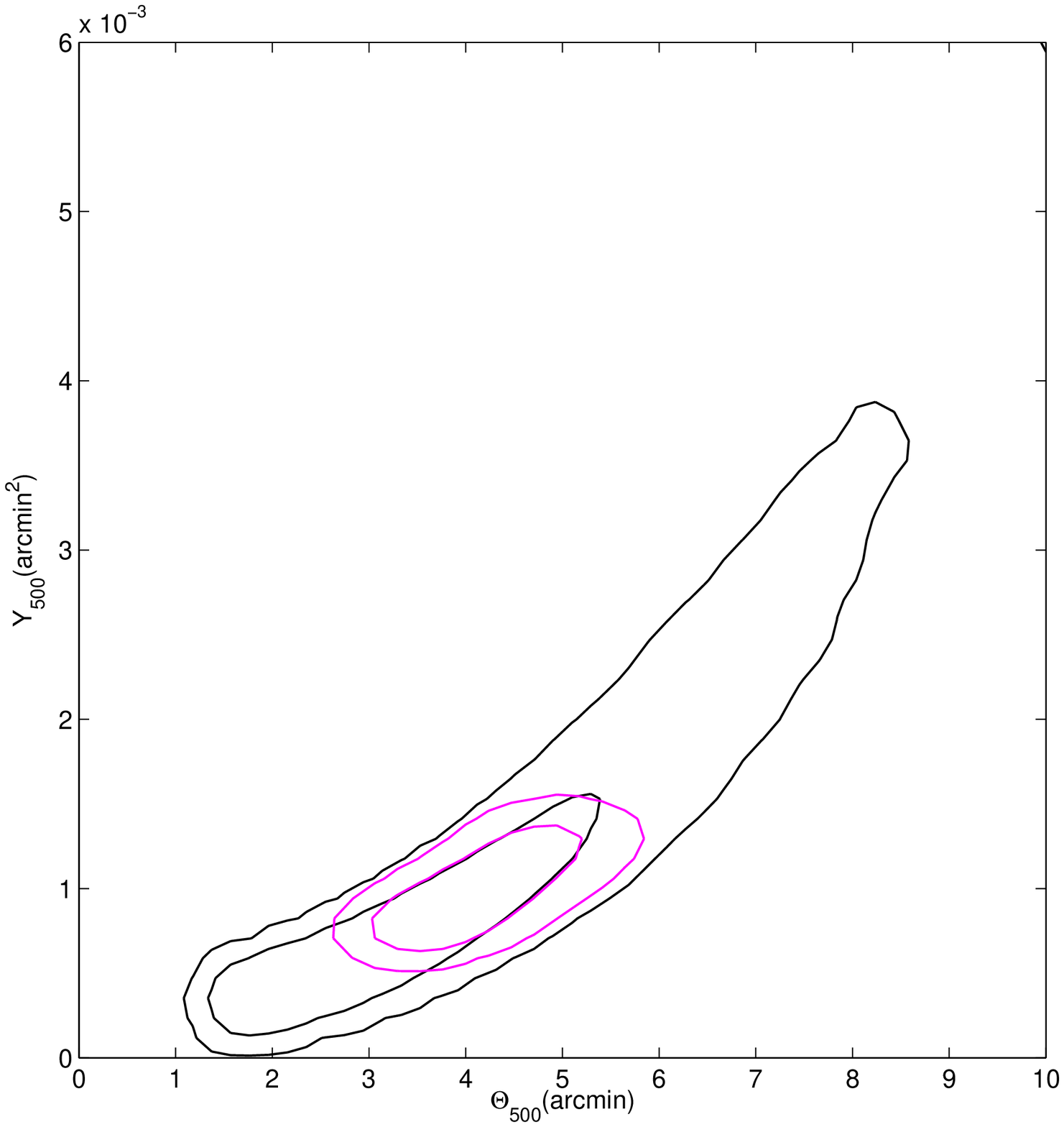}}\\
\subfigure[Left to right:   P190, P205, P351]{
\centering
\includegraphics[width=6.5cm, height=6.5cm,trim=60 48 60 48,clip= ,angle=0.]{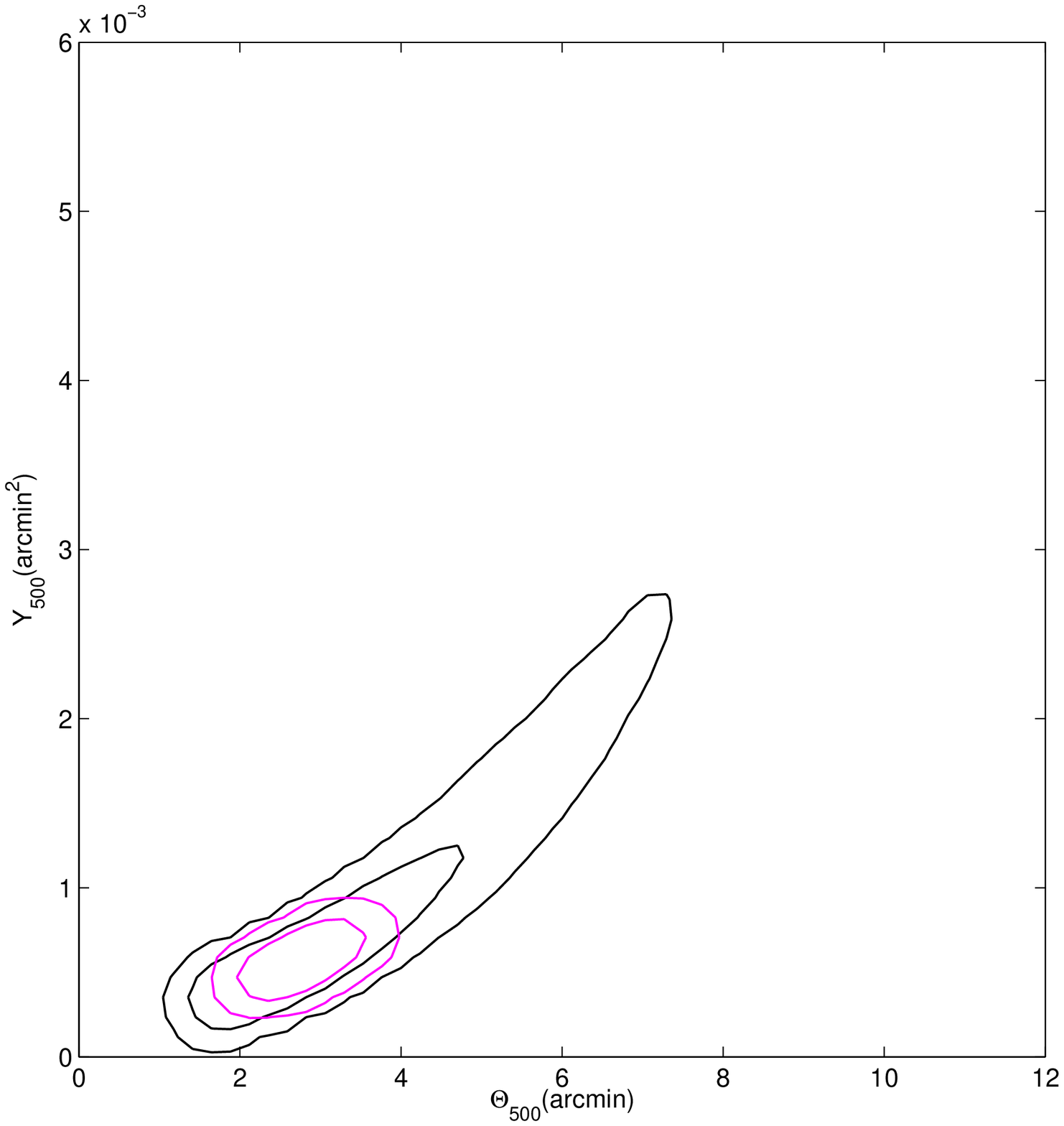}\hspace{0.05em}\includegraphics[width=6.5cm, height=6.5cm,trim=60 48 60 48,clip=,angle=0.]{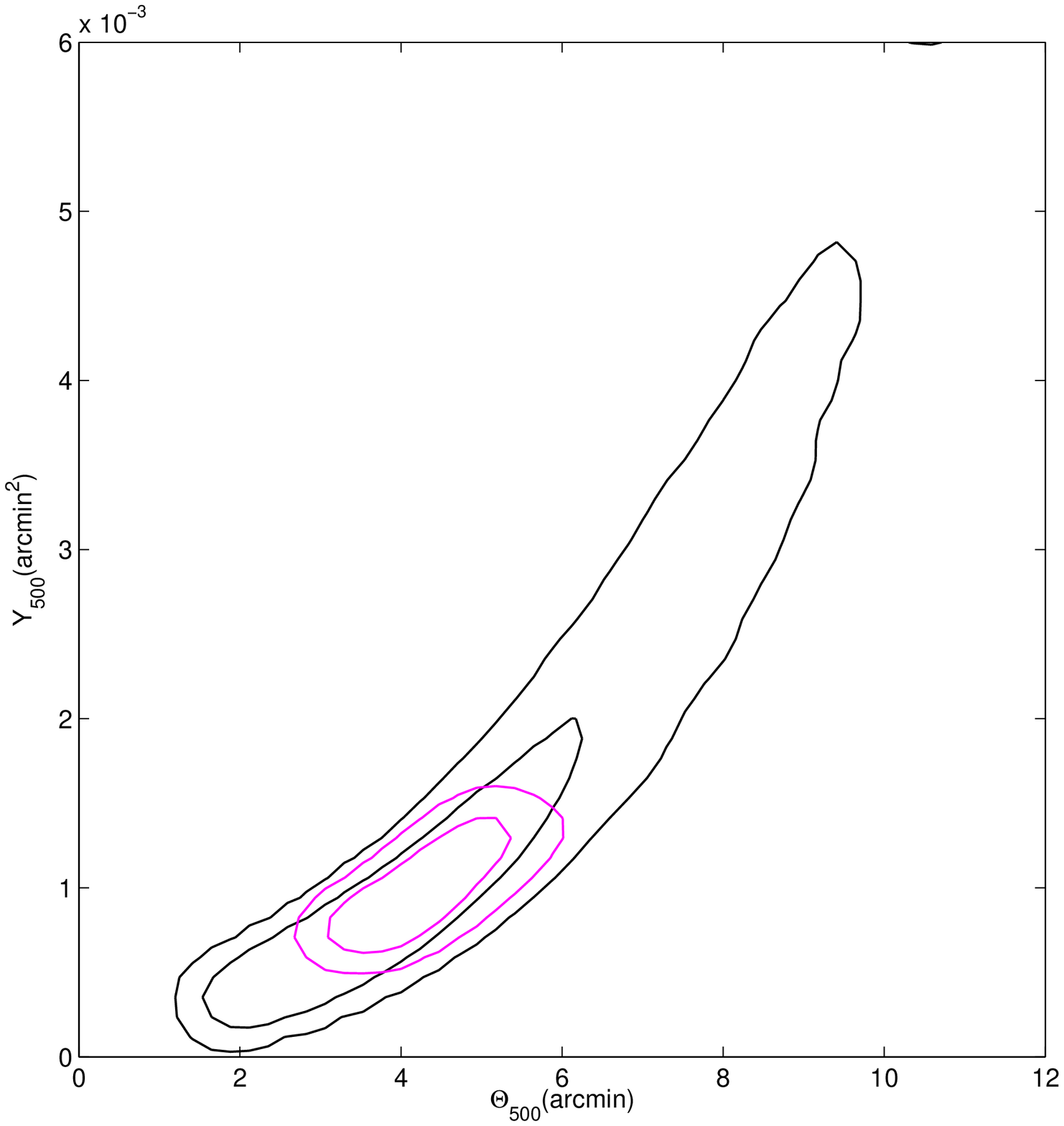}\hspace{0.05em}\includegraphics[width=6.5cm,height=6.5cm,trim=60 32 60 48,clip=,angle=0.]{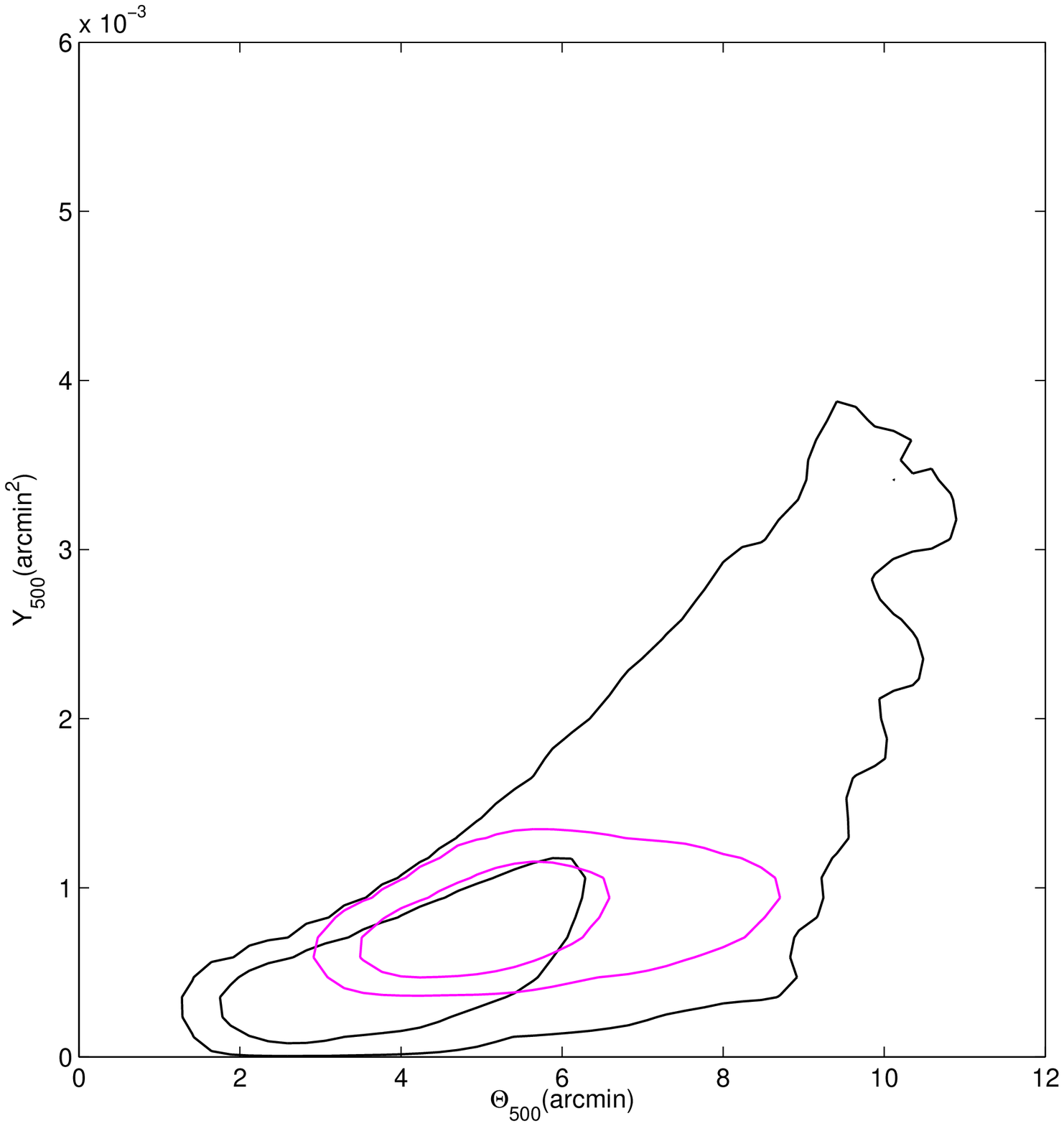}}\\
\caption{2D posterior distributions in the $Y_{500}-\theta_{500}$ plane for the CARMA-8 data alone in black contours and for the CARMA-8 data using a {\it{Planck}} prior on $Y_{500}$ (column 4 in Table \ref{tab:paramsY500}), in magenta contours, to which we refer to as the joint CARMA-{\it{Planck}} constraints. The y-axis is $Y_{500}$ in units of $\rm{arcmin}^2$ and the x-axis is $\theta_{500}$ in units of arcmin. The lower and upper limits of the y-axis are 0 and 0.006\,$\rm{arcmin}^2$, labelled in  steps of 0.001\,$\rm{arcmin}^2$ and for the x-axis they are 0 to 12 arcmin, labelled in steps of 2 arcmin. The inner and outer contours in each set indicate the areas enclosing the 68$\%$ and 95$\%$ of the probability distribution. For all the candidate clusters there is dramatic improvement in  $Y_{500}-\theta_{500}$ uncertainties when the CARMA and {\it{Planck}} data are jointly analyzed to yield cluster parameters.}
\label{fig:combi}
\end{center}
\end{figure*}

\begin{figure*}
\begin{center}
\centerline{\includegraphics[width=8cm,height=8cm,clip=,angle=0.]{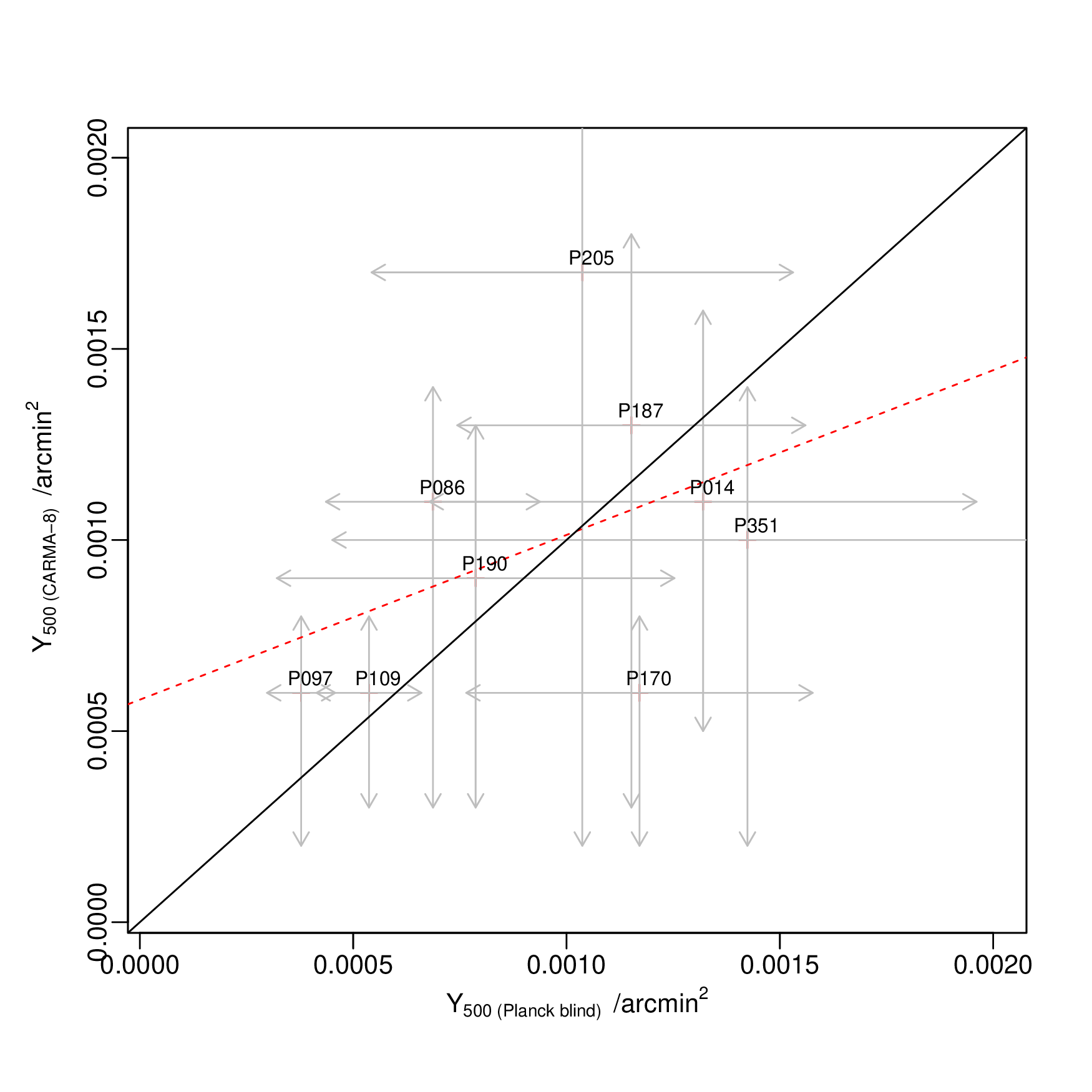}\qquad\includegraphics[width=8cm,height=8cm,clip=,angle=0.]{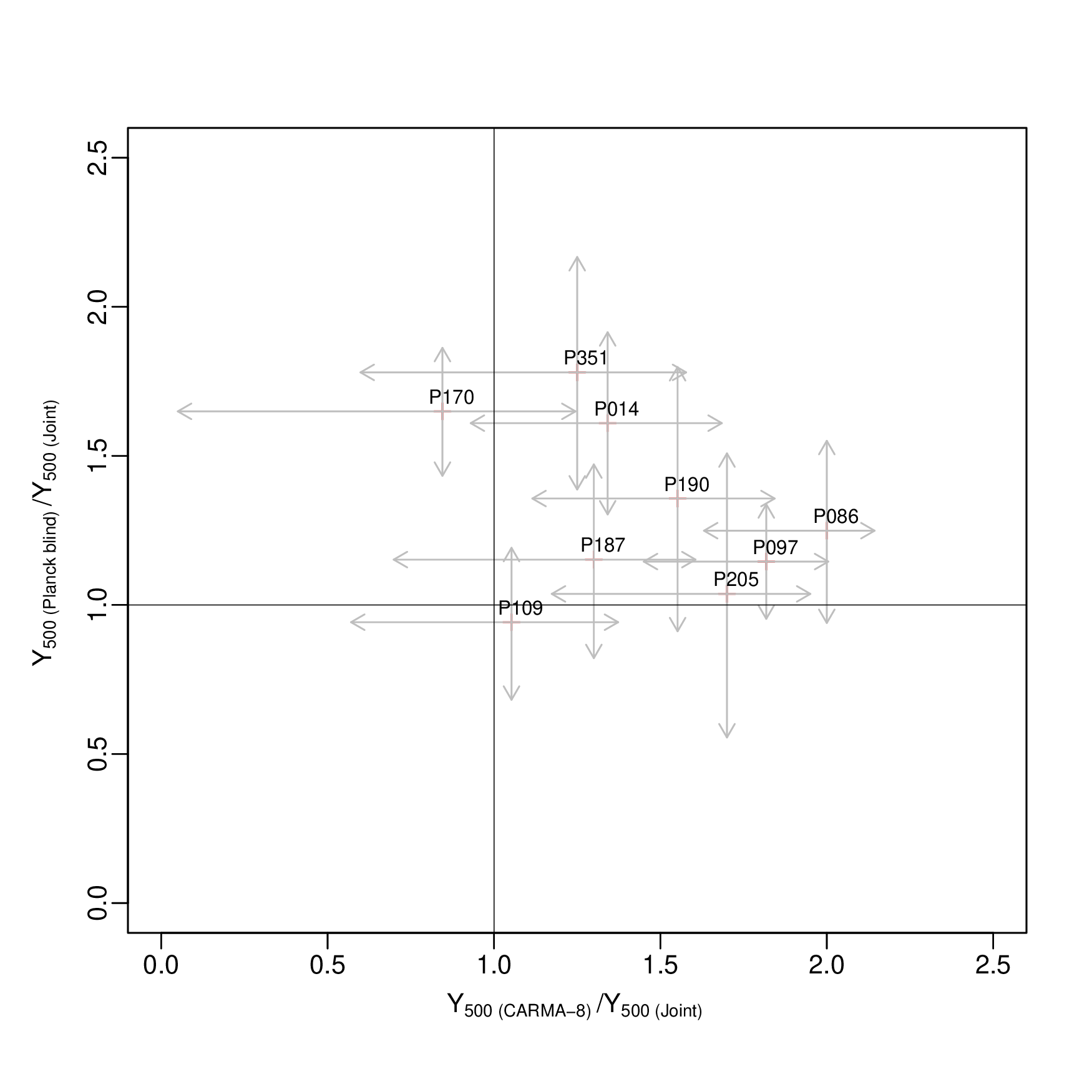}}
\caption{Left: $Y_{500}$ measured by CARMA-8 against the blind {\it{Planck}} $Y_{500}$. No strong bias indications are detected between the {\it{Planck}}, blind and CARMA-8 $Y_{500}$ results. The best-fit (red, dotted) and 1:1 (black, solid) lines are included in this plot. Right: ratio of the {\it{Planck}}, blind  $Y_{500}$ and the joint $Y_{500}$ values against the ratio of the CARMA-8 $Y_{500}$ and the joint $Y_{500}$. For $Y_{500}$, both the CARMA-8 and {\it{Planck}}, blind values are practically always higher than the joint values by as much as a factor of $2$. }
\label{fig:y500plots}
\end{center}
\end{figure*}

\begin{figure*}
\begin{center}
\centerline{\includegraphics[width=8cm,height=8cm,clip=,angle=0.]{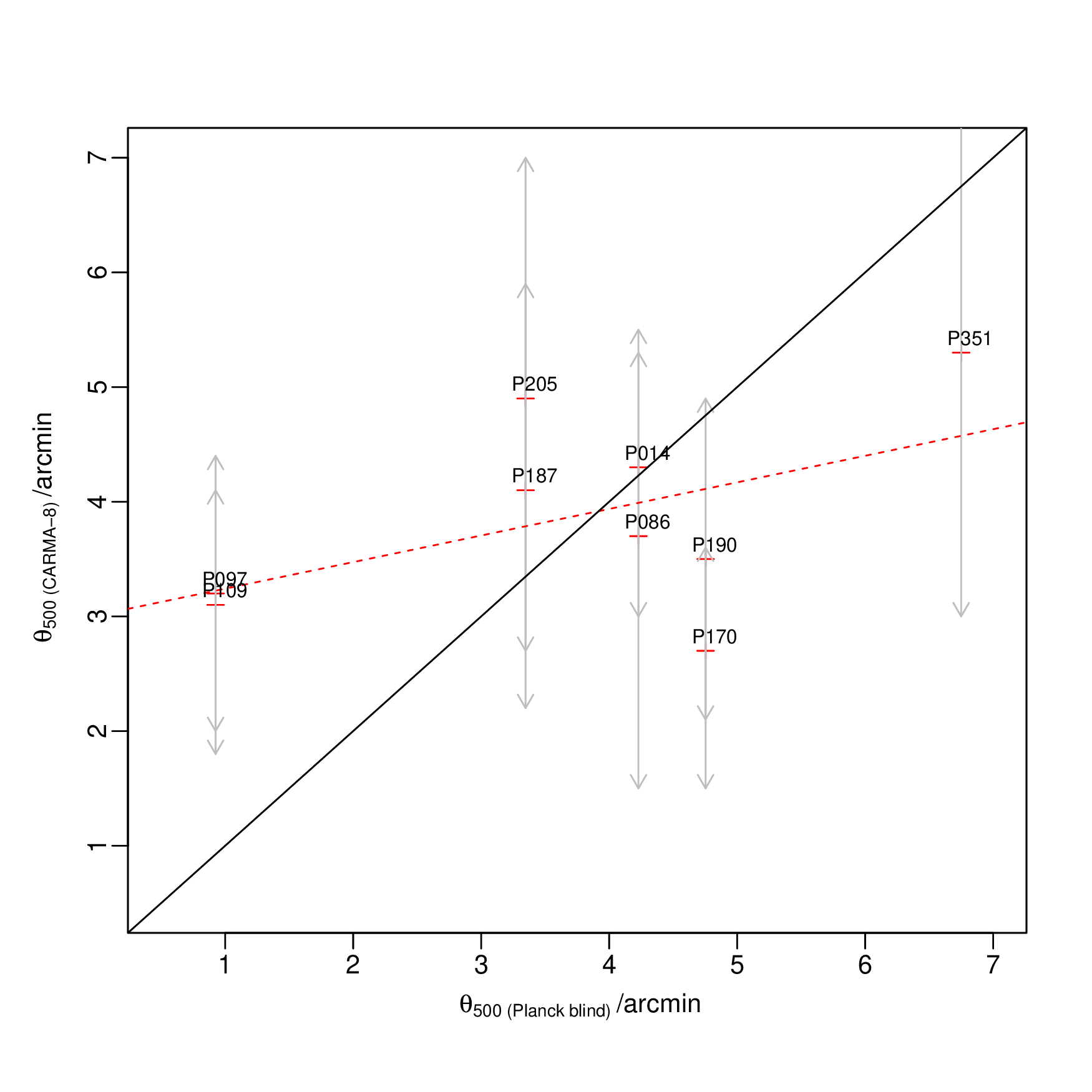}\qquad\includegraphics[width=8cm,height=8cm,clip=,angle=0.]{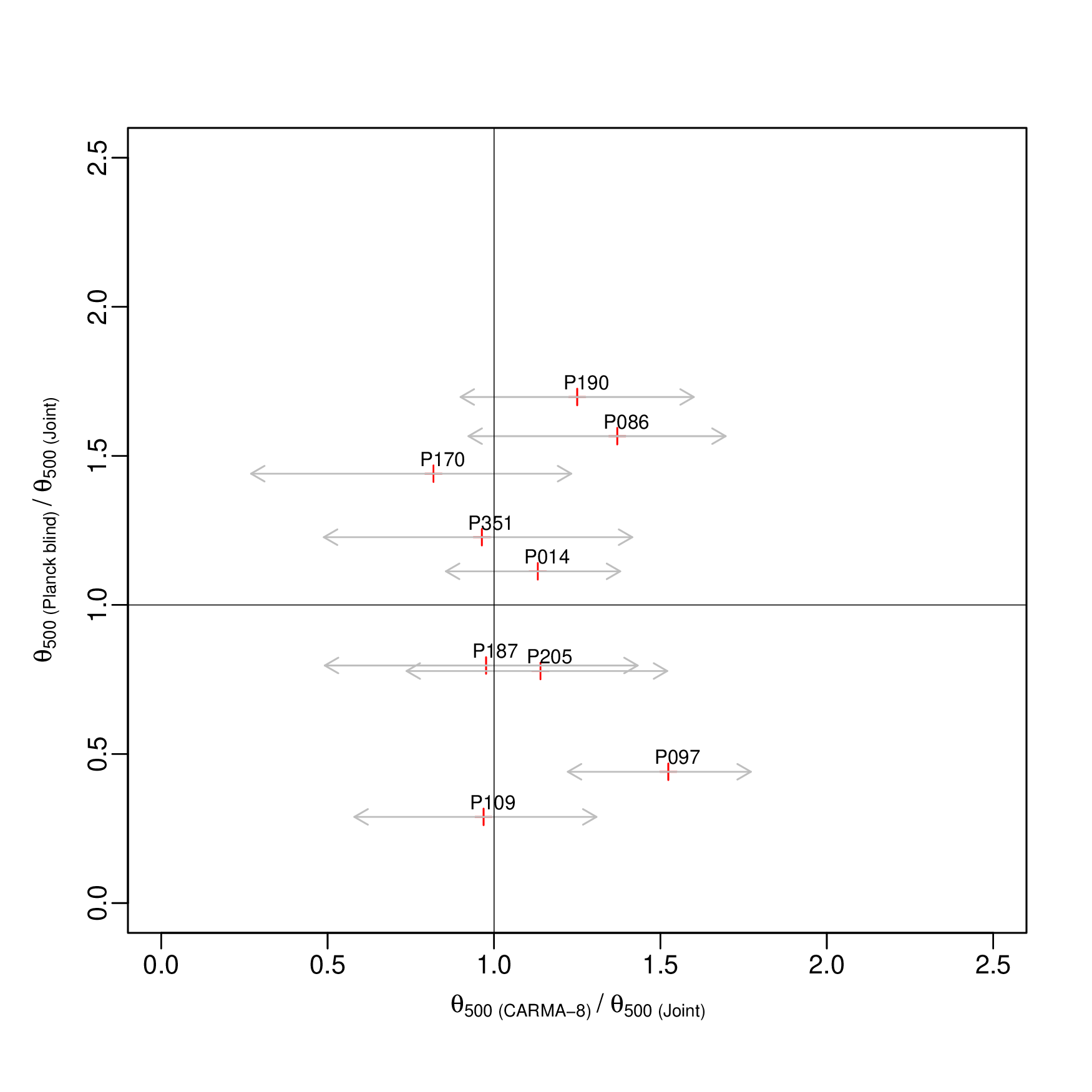}}
\caption{Left: $\theta_{500}$ measured by CARMA-8 against the blind {\it{Planck}} $\theta_{500}$. Uncertainties on the {\it{Planck}}, blind $\theta_{500}$ are not included as they are determined by the large FWHM of the {\it{Planck}}  beam and can range up to 10\arcmin. The best-fit (red, dotted) and 1:1 (black, solid) lines are included in this plot.  Right: Ratio of the blind {\it{Planck}} $\theta_{500}$ and the joint $\theta_{500}$ against the ratio of the CARMA-8 $\theta_{500}$ and the joint $\theta_{500}$. No strong systematic differences are seen between the different $\theta_{500}$ estimates. The higher resolution of the CARMA-8 data allows $\theta_{500}$ to be constrained much more strongly than using {\it{Planck}} data alone. }
\label{fig:theta500plots}
\end{center}
\end{figure*}

\subsection{The $Y_{500}$-$\theta_{500}$ Degeneracy}

In Figure \ref{fig:combi}, the 2D marginalized posterior distributions in the $Y_{500}-\theta_{500}$ plane for (i) the CARMA-8 data alone and (ii) the joint analysis of CARMA-8 and {\it{Planck}} data are displayed for each cluster candidate with an SZ detection in the CARMA-8 data. It is clear from these plots and from Table \ref{tab:paramsY500} that there is good overlap between the {\it{Planck}} and CARMA-8 derived cluster parameter space for all of the candidate clusters. The range of {\it{Planck}} $Y_{500}$ values after application of the $\theta_{500}$ priors from CARMA-8 are within the 68$\%$ contours for the CARMA-only analysis. While the {\it{Planck}} only range of $Y_{500, blind}$ values is wide, in combination with the $\theta_{500}$ constraints from the higher resolution CARMA-8 data, the $Y_{500}-\theta_{500}$ space is significantly reduced.
To explore this further, in Figure \ref{fig:y500plots} we have plotted: \\
 (i) the relation between $Y_{500}$ from CARMA-8 and from the {\it{Planck}}, blind analysis $Y_{500,{\rm{blind}}}$ (left panel) \\
 (ii) the ratio of $Y_{500}$ from the {\it{Planck}}, blind analysis and from the joint results against the ratio of $Y_{500}$ from fits  to CARMA-8 data and from the joint analysis (right panel). \\
 In Figure \ref{fig:theta500plots}, we present similar plots for $\theta_{500}$.
 Inspection of Figures \ref{fig:y500plots} and \ref{fig:theta500plots}, left panels, shows that the {\it{Planck}}, blind values for $Y_{500}$, as well as for $\theta_{500}$, appear unbiased with respect to those from CARMA-8.  As expected, due to {\it{Planck}}'s low angular resolution, the overall agreement between the {\it{Planck}}, blind measurements and those of CARMA-8 are much better for $Y_{500}$ than for $\theta_{500}$, with $\left< \Theta_{500, {\rm{CARMA-8}}}/\Theta_{500,{\rm{Planck\,\,blind}}} \right > =  1.5$, s.d.=1.1 and $\left < Y_{500, \rm{CARMA-8}}/ Y_{500, \rm{Planck\,\,blind}} \right> = 1.1$, s.d.=0.4. This good agreement in $Y_{500}$ is in contrast with recent results by \cite{linden2014}, who compare cluster-mass estimates derived by {\it{Planck}} and weak lensing data towards 22 clusters and find {\it{Planck}} masses are lower than the weak lensing masses typically by $\approx 30\%$. Such a bias would indeed alleviate the tension found in \citep{PlanckCosmos}, where the $95\%$ probability contours in the $\sigma_8-\Omega_{\rm{m}}$ plane derived from cluster data and from the CMB temperature power spectrum do not agree, when accounting for up to a $20\%$ bias from the assumption of hydrostatic equilibrium in the X-ray-based cluster-mass scaling relations. Since we find no signs of bias between the $Y_{500}$ measurements from {\it{Planck}} and CARMA-8, this might indicate the bias arises when comparing masses rather than $Y_{500}$, that is, from the choice of scaling relations used to estimate the cluster mass. Although, our cluster sample is relatively small and our parameter uncertainties are substantial other sources of bias in the SZ, X-ray and lensing measurements need to be investigated further with the same samples of objects rather than cluster samples selected in different ways and extending over different redshift ranges.
  
The right panel of Figure \ref{fig:y500plots} shows that, for all but two candidate clusters, the $Y_{500}$ measurements derived from the joint analysis are consistently lower than from the independent analysis of the {\it{Planck}}, blind and CARMA-8 datasets, sometimes by as much as a factor of 2. The average magnitude and range of the shift from the independent-to-joint $Y_{500}$ values are similar for the CARMA-8 and the {\it{Planck}}, blind analysis, with $\left <\frac{Y_{500, {\rm{blind}}}}{Y_{500, {\rm{Joint}}}}\right > = 1.3$, s.d.= 0.3, $\left <\frac{Y_{500, {\rm{CARMA-8}}}}{Y_{500, {\rm{Joint}}}}\right >=1.4 $, s.d. = 0.4.
In the case of $\theta_{500}$, Figure \ref{fig:theta500plots} (right), there is no systematic offset between the joint results and those from either CARMA-8 or {\it{Planck}}, blind. On average, the agreement between the joint and independent results appears to be good,  $\left <\frac{\Theta_{500, {\rm{blind}}}}{\Theta_{500, {\rm{Joint}}}}\right > =1.0$, s.d. = 0.5,  $\left <\frac{\Theta_{500, {\rm{CARMA-8}}}}{\Theta_{500, {\rm{Joint}}}}\right >=1.1$, s.d. = 0.2, but, in the case of the {\it{Planck}}, blind measurements, the large standard deviations are an indication of its poor resolution.

In Table \ref{tab:snrimp} we quantify the improvement in the constraints for $Y_{500}$ and $\theta_{500}$ derived from the independent analyses of {\it{Planck}} and CARMA-8 data with respect to the joint results. The largest improvement is seen for $\theta_{500}$ {\it{Planck}} blind, as this parameter is only weakly constrained by {\it{Planck}} data alone, with an associated uncertainty anywhere up to $\approx10\arcmin$. Yet, the advantage of a joint {\it{Planck}} and CARMA-8 analysis is very significant both for $\theta_{500}$ and $Y_{500}$ and for both the CARMA-8 and {\it{Planck}} results, with uncertainties dropping by $\gtrsim 400\%$. As mentioned in Section \ref{sec:joint}, the SZ measurements from {\it{Planck}} and CARMA-8 data are complementary as they probe different cluster scales at different resolutions. Moreover, since the $Y_{500}-\theta_{500}$ degeneracy for each data set have different orientations (an effect already reported in \citealt{PlanckAMI}), a joint analysis looking at the overlapping regions will result in a further reduction of parameter space.

\begin{figure*}
\begin{center}
\centerline{\includegraphics[width=9cm,height=9cm,clip=,angle=0.]{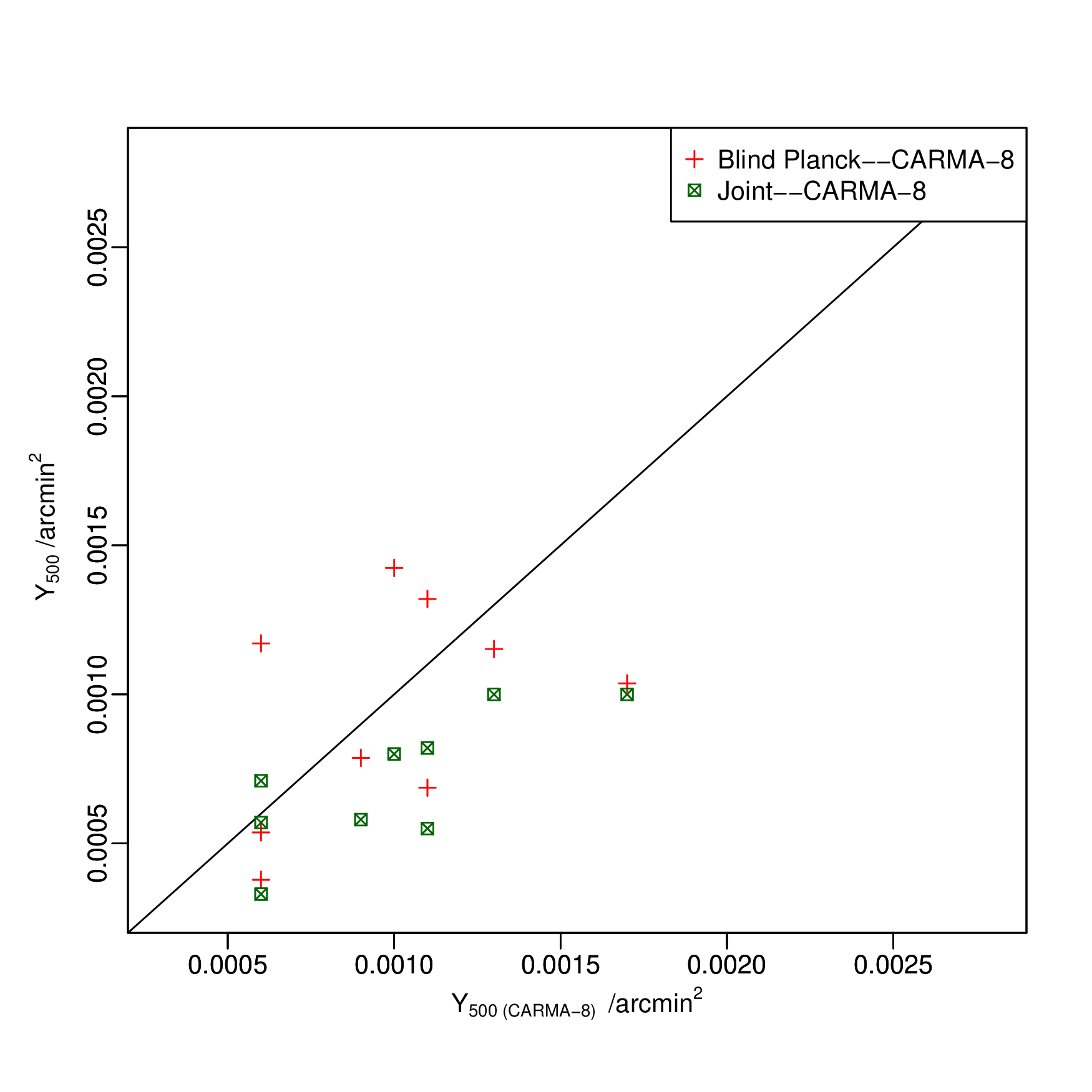}\quad\includegraphics[width=9cm,height=9cm,clip=,angle=0.]{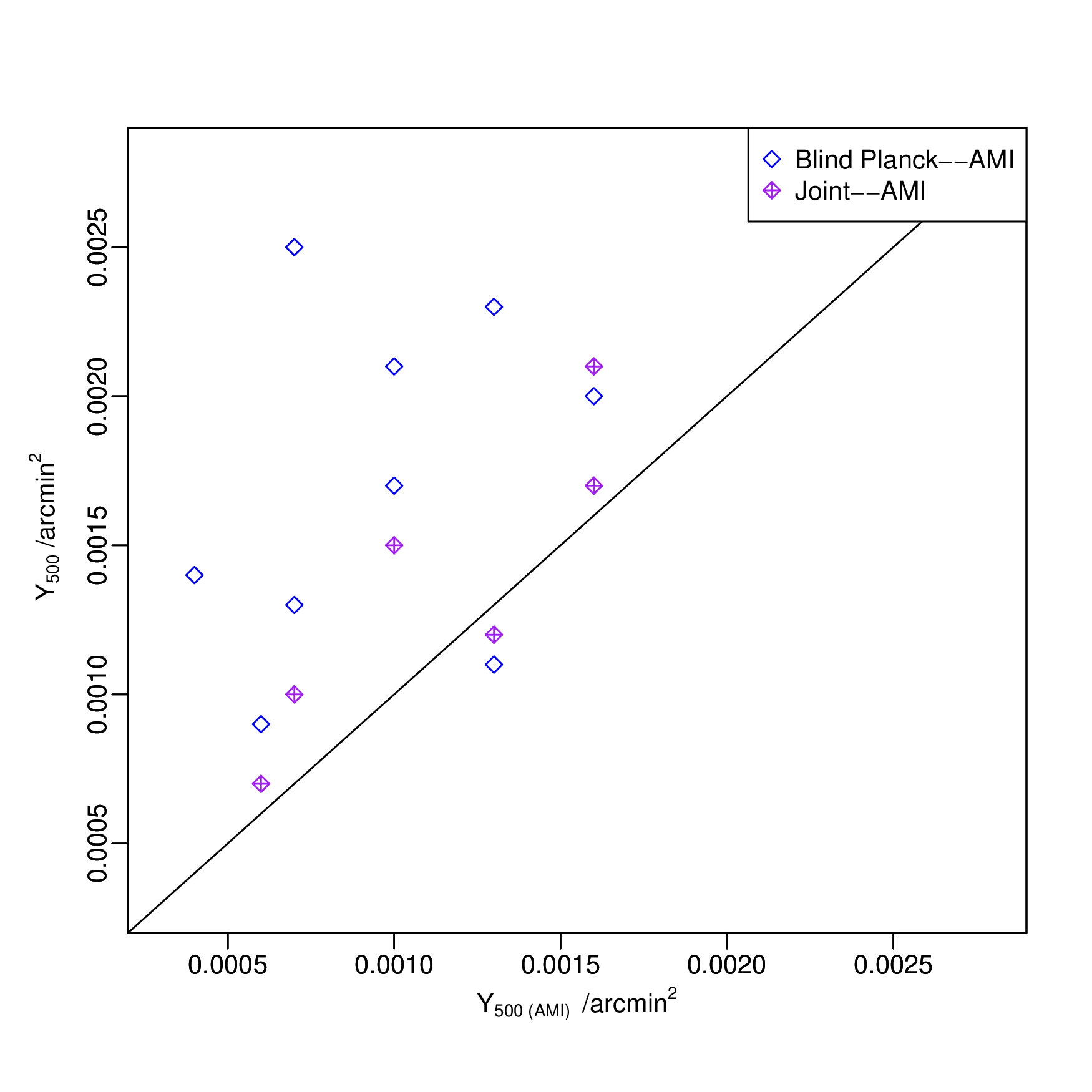}}
\caption{Left: Plot of $Y_{500}$ derived from two different analyses of {\sc{Planck}} data against $Y_{500}$ derived from CARMA-8 data. We plot two sets of $Y_{500}$ based on {\it{Planck}} measurements (i) using {\it{Planck}} data alone; these data points are referred to as `blind' and are shown in red crosses (ii)  using the range of $Y_{500}$ {\it{Planck}} (blind) values to constrain further the $Y_{500}$ values obtained in the analysis of CARMA-8 data; these are referred to as `joint' values and are shown in green crosses inside a square. In addition, we plot a $1:1$ line in solid black. For the joint $Y_{500}$ values, these are plotted on the y-axis, whilst keeping the x-axis as the CARMA-8 $Y_{500}$ values. Right: Plot of $Y_{500}$ from several analyses of {\sc{Planck}} data against $Y_{500}$ from AMI. We plot two sets of $Y_{500}$ based on {\it{Planck}} measurements (i) using {\it{Planck}} data alone; these data points are referred to as `blind' and are shown in blue diamonds (ii) using the range of $Y_{500}$ {\it{Planck}} (blind) values as a prior in the analysis of the AMI data; these are referred to as `joint' values and are shown in purple + signs inside a diamond. For the joint $Y_{500}$ values, these are plotted on the y-axis, whilst keeping the x-axis as the AMI $Y_{500}$ values. In addition, we plot a $1:1$ line in solid black. Comparison of the left and right plots shows the {\it{Planck}} (blind) measurements are generally in good agreement with CARMA-8, with no signs of systematic offsets between the two measurements. Calculating the {\it{Planck}} $Y_{500}$ using the CARMA-8 $\theta_{500}$ and position measurements, decreases the {\it{Planck}} $Y_{500}$ estimates, which become systematically slightly lower than the CARMA-8 estimates (Table \ref{tab:paramsY500}). This systematic difference is enhanced for the joint $Y_{500}$ values. For AMI, the {\it{Planck}} blind $Y_{500}$ values appear to be consistently higher than the AMI values for all but one cluster. While this difference narrows for the joint AMI-{\it{Planck}} $Y_{500}$ estimates, it is not resolved. Hence, the AMI $Y_{500}$ estimates are generally higher than those for {\it{Planck}}, irrespective of the choice of {\it{Planck}}-derived $Y_{500}$ while CARMA-8 $Y_{500}$ values are in good agreement with the (blind) {\it{Planck}} results, yet are generally higher when priors are applied to the CARMA-8 or {\it{Planck}} data.}
\label{fig:AMI}
\end{center}
\end{figure*}

\subsection{Comparison with the AMI-{\it{Planck}} Study}
\label{sec:AMI}

The Arcminute MicroKelvin Imager (\citealt{zwart2008}) has followed up $\approx 100$ {\it{Planck}}-detected systems, most of which are previously confirmed systems. Comparison of AMI and {\it{Planck}} results for 11 clusters in \citealt{PlanckAMI} showed that, as seen by {\it{Planck}}, clusters appear larger and brighter than by AMI. This result has now been confirmed in a larger upcoming study (\citealt{perrott2014}). In Figure  \ref{fig:AMI} we compare AMI and CARMA-8 values for $Y_{500}$ against the {\sc{Planck}} results. We note that the analysis pipeline for the processing of CARMA-8 data in this work is identical to that of AMI, allowing for a clean comparison between both studies.

 All eleven clusters in the AMI-{\sc{Planck}} study are known X-ray clusters (except for two) and are at $0.11 < z < 0.55$, with an average $z$ of 0.3 and a mean $\theta_{500}$ of 4.8. Our sample of cluster candidates is expected to have a mean (coarse) photometric redshift of $\approx 0.5$ (see Paper 1) and a mean $\theta_{500}$ of 3.9. The smaller angular extent of our sample of objects could be indicative that they are in fact at higher redshifts. 
 As pointed out in \cite{PlanckAMI}, the $Y_{500}$ measurements from {\it{Planck}} are systematically higher than those for AMI, $\left< Y_{500,  \rm{AMI}}/Y_{500, \rm{blind\,Planck}} \right> = 0.6$, s.d. = 0.3.  For the CARMA-8 results, on the other hand,  we find good agreement between the 
CARMA-8 and {\it{Planck}}-blind $Y_{500}$ values, $\left< Y_{500,  \rm{CARMA-8}}/Y_{500, \rm{blind\,Planck}} \right> = 1.1$, s.d. = 0.4, with no systematic offset between the two measurements.

 \cite{PlanckAMI} point to the use of a fixed gNFW profile as a likely source for systematic discrepancies between the AMI -{\it{Planck}} results and they plan to investigate changes to the results when a wider range of profiles are allowed in the fitting process. With relatively similar spatial coverage between the CARMA-8 and AMI interferometers, the impact of using a gNFW profile in the analysis of either data set should be comparable (for most cluster observations), though this will be investigated in detail in future work. The fact that, despite this, the CARMA-8 measurements are in good agreement with {\it{Planck}} could mean that either the CARMA-8 and {\it{Planck}} data are both biased-high due to systematics yet to be determined, or the AMI data are biased low, or a combination of both of these options. We stress that the analysis pipeline for deriving cluster parameters in this work and in \cite{PlanckAMI} is the same. Apart from the (typically) relatively small changes in the $uv$ sampling of both instruments, the major difference between the two datasets is that radio-source contamination at 16\,GHz tends to be much stronger than at 31\,GHz, since radio sources in this frequency range tend to have steep falling spectra and, hence, it tends to have a smaller impact on the CARMA-8 data. That said, AMI has a separate array of antennas designed to make a simultaneous, sensitive, high resolution map of the radio-source environment towards the cluster, in order to be able to detect and accurately model contaminating radio sources, even those with small flux densities ($\approx 350\mu$Jy/beam). 
 
A more likely alternative is the intrinsic heterogeneity in the different cluster samples. At low redshifts, when a cluster is compact relative to the AMI beam, heterogeneities in the cluster profiles get averaged out and the cluster integrated $Y_{500}$ values agree with those from {\it Planck}. However, if the cluster is spatially extended relative to the beam, a fraction of the SZ flux is missed and that results in an underestimate of the AMI SZ flux relative to the  {\it Planck} flux. Since our sample is at higher redshifts and appears to be less extended compared to the clusters presented in  \cite{PlanckAMI}, it is likely that profile heterogeneities are averaged out
 alleviating this effect seen in the low-z clusters. The prediction therefore, is that more distant, compact clusters that are followed up by AMI will show better agreement in $Y_{500}$ values with {\it Planck} although
 hints of a size-dependency to the agreement are already seen in \cite{PlanckAMI}.
 
 The results from this comparison between {\it{Planck}}, AMI and CARMA-8 SZ measurements draw further attention to the need to understand the nature of systematics in the data, in order to use accurate cluster-mass estimates for cosmological studies. To address this, we plan to analyze a sample of clusters observed by all three instruments in the future.

\section{Conclusions}

We have undertaken high (1-2\arcmin) spatial resolution 31 GHz observations with CARMA-8 of 19 {\it{Planck}}-discovered cluster candidates associated with significant overdensities of galaxies in the {\sc{WISE}} early data release ($\gtrsim1$\,galaxies/arcmin$^2$). The data reduction, cluster validation and photometric-redshift estimation were presented in a previous article (Paper 1; \citealt{rodriguez2014a}). In this work we used a Bayesian-analysis software package to analyze the CARMA-8 data. Firstly, we used the Bayesian evidence to compare models with and without a cluster SZ signal in the CARMA-8 data to determine that nine clusters are robust SZ detections and that two candidate clusters are most likely spurious. The data quality for the remaining targets was insufficient to confirm or rule out the presence of a cluster signal in the data.

Secondly, we analyzed the 9 CARMA-8 SZ detections with two cluster parameterizations. The first was based on a fixed-shape gNFW profile, following the model used in the analysis of {\it{Planck}} data (e.g., \citealt{ESZ}), to facilitate a comparative study. The second was based on a $\beta$ gas density profile that allows for the shape parameters to be fit. There is reasonable correspondence for the cluster characteristics derived from either parameterization, though there are some exceptions. In particular, we find that the volume-integrated brightness temperature within $\theta_{500}$ calculated using results from the $\beta$ profile does not correlate well with $Y_{500}$ from the gNFW parameterization for two systems. This suggests that differences in
the adopted profile can have a significant impact on the cluster-parameter constraints derived from CARMA-8 data. Indeed, radial brightness temperature profiles for individual clusters  obtained using the $\beta$ model results exhibit a level of heterogeneity distinguishable outside the uncertainty, with the degree of concentration for the profiles within $\theta_{500}$ and within 600\arcsec ($\approx$ the FWHM of the 100-GHz {\it{Planck}} beam) being between $\approx 1$ and $2.5$ times different.

Cluster-parameter constraints from the gNFW parameterization showed that, on average, the CARMA-8 SZ centroid is displaced from that of {\it{Planck}} by $\approx1.5\arcmin$. Overall, we find that our systems have relatively small $\theta_{500}$ estimates, with a mean value of 3.9$\arcmin$. This is a factor of two smaller than for the MCXC clusters, whose mean redshift is $0.18$. This provides further, tentative, evidence to support the photometric-redshift estimates from Paper 1, which expected 
the sample to have a mean $z$ of $\approx 0.5$. Using Keck/MOSFIRE Y-band spectroscopy, we were able to confirm the redshift of a likely galaxy-member of one of our cluster candidates, P097, to be at $z=0.565$.

We analyzed data towards P190, a representative candidate cluster for the sample, with a cluster parameterization that samples from the cluster mass. This parameterization requires $z$ as an input. We ran it from a $z$ of 0.1 to 1 in steps of 0.2. Beyond the $z>0.3$ regime, the dependence of mass on $z$ is very mild (with the mass remaining unchanged to within 1 significant figure). Our estimate for $M_{500}$ at the expected average $z$ of our sample, 0.5 (Paper 1), is $0.8\pm0.2\times10^{15}\,M_{\odot}$.

We compared the {\it{Planck}} (blind) and CARMA-8 measurements for $Y_{500}$ and $\theta_{500}$. Both sets of results appear to be unbiased and in excellent agreement, with $\left < Y_{500, \rm{CARMA-8}}/ Y_{500, \rm{blind, Planck}} \right> = 1.1$, s.d.=0.4. and $\left< \Theta_{500, {\rm{CARMA-8}}}/\Theta_{500,{\rm{blind, Planck}}} \right > =  1.5$, s.d.=1.1, whose larger difference is a result of the poor spatial resolution of {\it{Planck}}. This is in contrast with the results from a similar study between AMI and {\it{Planck}} that reported systematic differences between these parameters for the two instruments, with {\it{Planck}} characterizing the clusters as larger (typically by $\approx 20\%$) and brighter (by $\approx 35\%$ on average) than AMI. However, it should be emphasized that the clusters studied in this paper are on average at higher redshift
and more compact than those in the AMI-{\it Planck} joint analysis.

Our results, and those from AMI, seem to not support the recent results by \cite{linden2014}, who find {\it{Planck}} masses to be biased-low by $\approx 30\%$ with respect to weak-lensing masses and, potentially, with \cite{PlanckCosmos} who consider the possibility of {\it{Planck}} masses to be biased-low by $20-40\%$ to explain inconsistencies in their results from cluster data and the CMB temperature power spectrum. The good agreement between {\it{Planck}} and CARMA-8 $Y_{500}$ measurements, in contrast with the large differences in the {\it{Planck}} masses derived from X-ray-based scaling relations, could be an indication that the origin of the discrepancy lies primarily in the choice of scalings and in the heterogeneity in cluster profiles with increasing redshift. However, our sample size is small and our uncertainties substantial. A large multi-frequency study of clusters, which included {\it{Planck}} and a high-resolution SZ experiment, like CARMA-8 or AMI, to check for systematics in $Y_{500}$, as well as lensing and X-ray data to investigate differences in cluster-mass estimates, would be beneficial to fully address this. 

We exploited the complementarity of the {\it{Planck}} and CARMA-8 datasets---the former measures the entire cluster flux directly unlike the latter, which can, on the other hand, constrain $\theta_{500}$---to reduce the size of the $Y_{500}-\theta_{500}$ degeneracy by applying a {\it{Planck}} prior on the $Y_{500}$ obtained from CARMA-8 alone. We show how this joint analysis reduces uncertainties in $Y_{500}$ derived from the {\it{Planck}} and CARMA-8 data individually by more than a factor of $\gtrsim 5$.  

In this article and in its companion paper (\citealt{rodriguez2014a}), we have demonstrated (1) an interesting technique for the selection of massive clusters at intermediate $z\gtrsim 0.5$ redshifts by cross-correlating {\it{Planck}} data with WISE and other data (2) a powerful method for breaking degeneracies in the $Y_{500} ({\rm{flux}})-\theta_{500} ({\rm{size}})$ plane and thus greatly improving constraints in these parameters.

\section{Acknowledgements}

We thank the anonymous referee for a careful reading of the manuscript and would also like to acknowledge helpful discussions with Y. Perrott and A. Lasenby.
We thank the staff of the Owens Valley Radio observatory and CARMA for their outstanding support; in particular, we would like to thank John Carpenter. Support for CARMA construction was derived from the Gordon and Betty Moore Foundation, the Kenneth T. and Eileen L. Norris Foundation, the James S. McDonnell Foundation, the Associates of the California Institute of Technology, the University of Chicago, the states of California, Illinois, and Maryland, and the National Science Foundation. CARMA development and operations were supported by the National Science Foundation under a cooperative agreement, and by the CARMA partner universities. The {\it{Planck}} results used in this study are based on observations obtained with the {\it{Planck}} satellite (http://www.esa.int/Planck), an ESA science mission with instruments and contributions directly funded by ESA Member States, NASA, and Canada. We are very grateful to the AMI Collaboration for allowing us to use their software and analysis techniques in this work and for useful discussions. A small amount of the data presented herein were obtained at the W.M. Keck Observatory, which is operated as a scientific partnership among the California Institute of Technology, the University of California and the National Aeronautics and Space Administration. The Observatory was made possible by the generous financial support of the W.M. Keck Foundation.

\end{document}